\documentclass{iopart}
\usepackage[left=2cm,top=2cm,right=3cm]{geometry} 
\usepackage{graphicx}
\usepackage{psfrag}

\usepackage{graphics,graphicx}
\usepackage{dcolumn}
\usepackage{iopams}
\usepackage[usenames]{color}
\usepackage[breaklinks=false,bookmarksnumbered]{hyperref}
\usepackage[T1]{fontenc}

\newcommand{\mc}{\mathcal}
\newcommand{\diff}[2]{\frac{{\rm d}#1}{{\rm d}#2}}
\newcommand{\diffl}[2]{{\rm d}#1/{\rm d}#2}
\newcommand{\scp}[2]{\left(\vct{#1}\cdot\vct{#2}\right)}
\newcommand{\crp}[2]{\left(\vct{#1}\times\vct{#2}\right)}

\newcommand{\pb}[2]{\left[#1,#2\right]}
\newcommand{\unsc}[1]   {\hat { \vct{#1}} }

\newcommand{\markscal}[1]{\left[  {#1} \right]}

\def\vct#1{\mathbf{#1}}
\newcommand{\vek}[1]{\boldsymbol{#1}}

\def\eanl{\\}
\def\neanl{\nonumber\\}

\def\sm{s_{-}}
\def\sp{s_{+}}
\def\Sone{S_{\rm{1}}}
\def\Stwo{S_{\rm{2}}}
\def\Pone{p}

\def\nunit{n_{\rm{12}}}
\def\mean{\ell}
\newcommand{\HAM}[2]{	{\rm H}_{\rm {#1}}^{\rm {#2}}}

\newcommand{\vPone}{\vct{\Pone}}
\newcommand{\vnunit}{\vct{\nunit}}
\newcommand{\vSone}{\vct{\Sone}}
\newcommand{\vStwo}{\vct{\Stwo}}
\newcommand{\vAng}{\vct{\ang}}

\newcommand{\vex}{\vct{e_x}}
\newcommand{\vey}{\vct{e_y}}
\newcommand{\vez}{\vct{e_z}}

\newcommand{\ver}{\vek{e_r}}
\newcommand{\vephi}{\vek{e_\phi}}
\newcommand{\vetheta}{\vek{e_\theta}}

\newcommand{\pN}{PN}
\newcommand{\aso}{\alpha_{\rm{so}}}
\newcommand{\asq}{\alpha_{\rm{s^{2}}}}
\newcommand{\ass}{\alpha_{\rm{s}_{1}\rm{s}_{2}}}

\def\ang{h} 
\def\epso{\epsilon} 
\def\spino{\delta} 
\def\rel{r} 

\newcommand{\dt}{( 1-e_t \, \cos u )}

\newcommand{\OTS}{\sqrt{1-e_t^2}}

\newcommand{\omfeta}{\sqrt{1-4\eta}}

\newcommand{\OutSCNn}{\scp{N}{\nunit}}
\newcommand{\OutSCNv}{\scp{N}{v}}

\newcommand{\SPAT}[3]{\left[#1 \cdot \left(#2 \times #3\right) \right]}

\newcommand{\OutXnn}{	{\cal P}^{(\times)}_{nn}}
\newcommand{\OutXvv}{	{\cal P}^{(\times)}_{vv}}
\newcommand{\OutXnv}{	{\cal P}^{(\times)}_{nv}}

\newcommand{\OutPnn}{	{\cal P}^{(+)}_{nn}}
\newcommand{\OutPvv}{	{\cal P}^{(+)}_{vv}}
\newcommand{\OutPnv}{	{\cal P}^{(+)}_{nv}}

\newcommand{\OutXPnn}{	{\cal P}^{(\times,+)}_{nn}}
\newcommand{\OutXPvv}{	{\cal P}^{(\times,+)}_{vv}}
\newcommand{\OutXPnv}{	{\cal P}^{(\times,+)}_{nv}}

\newcommand{\SCNn}{				\scp{\vct{N}}{\vnunit}}

\newcommand{\DeltaCrossn}[1]{			\crp{\vct{\Delta}}{\vnunit}^{#1}}
\newcommand{\DeltaCrossv}[1]{			\crp{\vct{\Delta}}{\vct{v}}^{#1}}
\newcommand{\DeltaCrossN}[1]{			\crp{\vct{\Delta}}{\vct{N}}^{#1}}
\newcommand{\SCrossn}[1]{			\crp{\vct{S}}{\vnunit}^{#1}}
\newcommand{\SCrossv}[1]{			\crp{\vct{S}}{\vct{v}}^{#1}}
\newcommand{\SCrossN}[1]{			\crp{\vct{S}}{\vct{N}}^{#1}}

\newcommand{\SCSnCrossv}{			\SPAT{\vct{S}}{\vnunit}{\vct{v}}		}
\newcommand{\SCvSCrossn}{			\SPAT{\vct{v}}{\vct{S}}{\vnunit}		}
\newcommand{\SCNSCrossv}{			\SPAT{\vct{N}}{\vct{S}}{\vct{v}}	}
\newcommand{\SCNSCrossn}{			\SPAT{\vct{N}}{\vct{S}}{\vnunit}	}

 \newcommand{\SCProjXPnDeltaCrossn}{	\markscal{	{\cal P}^{(\times, +)}_{ij}	\DeltaCrossn{i}	\nunit^{j}	}	}
 \newcommand{\SCProjXPnDeltaCrossv}{	\markscal{	{\cal P}^{(\times, +)}_{ij}	\DeltaCrossv{i}	\nunit^{j}	}	}

\newcommand{\SCProjXPvSCrossn}{		\markscal{	{\cal P}^{(\times, +)}_{ij}	\SCrossn{i}		v^{j}		}	}
 \newcommand{\SCProjXPnSCrossn}{		\markscal{	{\cal P}^{(\times, +)}_{ij}	\SCrossn{i}		\nunit^{j}	}	}
 \newcommand{\SCProjXPvSCrossv}{		\markscal{	{\cal P}^{(\times, +)}_{ij}	\SCrossv{i}		v^{j}		}	}
\newcommand{\SCProjXPnSCrossv}{		\markscal{	{\cal P}^{(\times, +)}_{ij}	\SCrossv{i}		\nunit^{j}	}	}
\newcommand{\SCProjXPvSCrossN}{		\markscal{	{\cal P}^{(\times, +)}_{ij}	\SCrossN{i}		v^{j}		}	}

\newcommand{\SCProjXPnSCrossN}{		\markscal{	{\cal P}^{(\times, +)}_{ij}	\SCrossN{i}	\nunit^{j}		}	}
\newcommand{\SCProjXPnDeltaCrossN}{	\markscal{	{\cal P}^{(\times, +)}_{ij}	\DeltaCrossN{i}	\nunit^{j}	}	}

\newcommand{\SCProj}[3]{ {\cal P}^{(#1)}_{ij} \,	{#2}{#3}}

\newcommand{\htt}[1]{h^{{\rm TT}}_{#1}}
\newcommand{\ttproj}[2]{{\cal P}^{{\rm TT}#1}_{#2}}
\newcommand{\pproj}[2]{{\cal P}^{(#1)}_{#2}}

\newcommand{\xii}[3]{\xi^{(#1)\,{\rm #2}}_{#3}}
\newcommand{\xip}[3]{\xi^{(#1)\,{\rm #2}}_{#3}}

\usepackage[T1]{fontenc}

\newcommand{\vvelocity}{\vct{v}}

\begin{document}

\title[Motion and gravitational wave forms of eccentric spinning compact binaries]
{
Motion and gravitational wave forms of eccentric compact binaries with orbital-angular-momentum-aligned spins under next-to-leading order
 in spin-orbit and leading-order in spin(1)-spin(2) and spin-squared coupling
}
\author{M Tessmer, J Hartung, and G Sch\"afer}
\address{Theoretisch-Physikalisches Institut, Friedrich-Schiller-Universit\"
		at Jena, Max-Wien-Platz 1, 07743 Jena, Germany}
\ead{\mailto{m.tessmer@uni-jena.de}}
\date{\today}

\begin{abstract}
A quasi-Keplerian parameterisation for the solutions of second
post-Newtonian (\pN) accurate equations of motion for spinning
compact binaries is obtained including leading order spin-spin
and next-to-leading order spin-orbit interactions.
Rotational deformation of the  compact objects is incorporated.
For arbitrary mass ratios the spin
orientations are taken to be parallel or anti-parallel to the orbital angular
momentum vector.
The emitted gravitational wave forms are given in analytic form up to 2\pN\ point particle,
1.5\pN\ spin-orbit and 1\pN\ spin-spin contributions,
whereby the spins are assumed to be of 0PN order.

\end{abstract}
\maketitle
\section{Introduction}

Inspiralling and merging neutron star (NS) and/or black hole (BH) binaries
are promising sources for continuous gravitational waves (GW).
Ground-based laser interferometers as e.g.
LIGO, VIRGO, and GEO are already searching for
those astrophysical sources \cite{Rowan:Hough:2000}.
For a successful search with the help
of {\em matched filtering} of the emitted GW signals, one needs
a detailed knowledge of the orbital dynamics.
	Spin effects of higher order
	were discussed in \cite{Faye:Blanchet:Buonanno:2006, Blanchet:Buonanno:Faye:2006,Blanchet:Buonanno:Faye:2006:err,Blanchet:Buonanno:Faye:2006:err:2,Arun:Buonanno:Faye:Ochsner:2009}
	for the inspiral of compact binaries
	were the orbits were assumed to be quasi-circular.
	A recent publication \cite{Hannam:Husa:Brugmann:Gopakumar:2008}
	gave a numerical insight into the evolution of binary systems
	having spins that are parallel to the orbital angular momentum
	and evolving in quasi-circular orbits.

Because there are many physical degrees of freedom involved,
it is computationally desirable to have an analytical
description, especially for interferometers working in
the early inspiral phase, where numerical relativity
currently fails to produce hundreds of orbital cycles.
For non-spinning compact binaries, the post-Newtonian (\pN) expansion 
in the near-zone has been
carried out through 3.5\pN\ order \cite{Jaranowski:Schafer:1997} and 3.5\pN\ accurate inspiral
templates have been established for circular orbits \cite{Blanchet:Faye:Iyer:Joguet:2002,Blanchet:Faye:Iyer:Joguet:2002:err}.
For numerical performances of these templates see \cite{Ajith:Babak:Chen:others:2008,Ajith:Babak:Chen:others:2008:err}.
Observations lead to the assumption that many astrophysical
objects carry a non-negligible {\em spin},
such that the effect of spin angular momentum cannot be
ignored for detailed data analysis.
The problem of spins in General Relativity (GR) was first discussed in
\cite{Mathisson:1937, Papapetrou:1951,Corinaldesi:Papapetrou:1951}
and considerable further developments were made
in the 1970s \cite{Barker:OConnell:1970, Barker:OConnell:1975, DEath:1975, Barker:OConnell:1979},
and in recent years as well.
Apostolatos \cite{Apostolatos:1995} showed in his analysis of
{\em simple precession} for ``circular'' orbits and spinning self-gravitating sources
that the form of the GW signal is affected. The amount of the energy radiated by the
binary system with spin has been determined by \cite{Kidder:Will:Wiseman:1993}.

Therefore, we want to implement recent breakthroughs in dynamical relativity of
spinning compact binaries into a useful prescription for data analysis applications.
Our aim is to connect the following items.

\begin{enumerate}
 \item The ``standard'' procedure to compute the evolution
of eccentric orbits from the Hamilton equations of motion (EOM).
For eccentric orbits, but neglecting spin effects, Damour
and Deruelle \cite{Damour:Deruelle:1985} presented a phasing at 1\pN\,
employing conchoidal transformations to reduce the structure
for the radial motion.
Later publications \cite{Schafer:Wex:1993,Schafer:Wex:1993:err,Memmesheimer:Gopakumar:Schafer:2004}
used Hamilton EOM instead of Lagrange ones and
employed a more general scheme for a solution to conservative 3\pN\
dynamics without spin.
\item
The 2\pN\ point particle (PP), next-to-leading order spin-orbit (NLO-SO)
and leading-order spin(1)-spin(2) (LO-S$_1$S$_2$) and spin(1)-spin(1)  (LO-S$^2$) interaction contributions.
\end{enumerate}
As a starting point, we assume (anti--) aligned spin and orbital angular
momentum vectors
for an estimation of the effects.
It is interesting to analyse this matter system configuration because numerical results
of a recent publication indicate that maximum equal-spins aligned with the
orbital angular momentum lead to
observable volume of up to $\sim$ 30 times larger than the corresponding binaries with the spins
anti-aligned to the orbital angular momentum \cite{Reisswig:Husa:Rezzolla:Dorband:Pollney:Seiler:2009}. 
From Figure 10 in \cite{Reisswig:Husa:Rezzolla:Dorband:Pollney:Seiler:2009},
one can also find an observable volume of those binaries up to
$\sim$ 8 times larger compared to non-spinning binaries.
These authors conclude that those systems are among the most efficient GW sources
in the universe.
In another recent publication \cite{Bogdanovic:Reynolds:Miller:2007} it can be found
that in gas-rich environments the spins of two black holes can align with the larger scale accretion disc on
a timescale that is short as $1\%$ of the accretion time. Due to the model of those authors, having two
black holes interacting independently with an accretion disc, their spins tend to be aligned with each other
and with the orbital angular momentum more or less depending on the model parameters.

We work only with the conservative Hamiltonian for the time being, and
restrict our attention to terms up to 2.5\pN\ order overall, assuming maximally 
spinning holes. This means neglecting both the well-known 3\pN\ PP contributions, 
and the NLO-$S_1 S_2$ \cite{Steinhoff:Hergt:Schafer:2008:2}, as well as the NLO-$S^2$ contributions, 
which have recently been derived for general compact binaries \cite{Hergt:Steinhoff:Schafer:2010:1}. 
This latter publication came out at a late stage in our calculations, but it should be a straightforward 
task to include these terms in a future publication.


	If the objects are slowly rotating, the considered leading-order spin-squared contributions
	are shifted to 3\pN\ order and, for consistency, the 3\pN\ PP Hamiltonian has to be included.
	The 3\pN\ PP contributions to the orbital elements are available in the literature
	\cite{Memmesheimer:Gopakumar:Schafer:2004} and simply have to be added to what we
	are going to present in this paper.
	Anyway, this work is consistently worked out to all terms up to 2.5\pN, having
	maximal rotation or not and will list all results in the spins which are counted of 0\pN\ order.

%

The paper is organised as follows:
\Sref{sec::Hamiltonian} summarises and discusses the Hamiltonian terms 
we want to include in our prescription.
\Sref{sec::ConservationOfParallel} investigates the conservation of
initial spin and orbital angular momentum alignment conditions.
In \Sref{sec::KeplerParam}, we briefly summarise the
Keplerian parameterisation for Newtonian orbital dynamics
and outline the generalisation to higher \pN\ accurate dynamics.
The solution of the Hamilton EOM is the subject of \Sref{sec::QKPSpin}.
\Sref{sec::Summary} summarises all important results of our procedure.
In \Sref{sec::Gravwave} we give some formulae for the polarisations of the
gravitational waves which are emitted by the system.
Calculations were mostly done with {\em Mathematica} and {\em xTensor}
(see \cite{MartinGarcia:2008,MartinGarcia:2002} and references therein).

\section{Spin and orbital dynamics}
\label{sec::Hamiltonian}

In the following sections, the dynamics of spinning compact binaries is
investigated, where the SO contributions are restricted to NLO
and the  ${\rm S}_1 {\rm S}_2$ and ${\rm S}^2$ to LO.
The PP contributions are cut off after the 2\pN\ terms.
The Hamiltonian associated therewith reads 
\begin{eqnarray}
\label{eq:dimH}
\hat {\rm H}
(\unsc{x}_{\rm 1}, \unsc{x}_{\rm 2}, \unsc{p}_{\rm 1}, \unsc{p}_{\rm 2}, \unsc{S}_{\rm 1}, \unsc{S}_{\rm 2})
= ~
&
   \hat \HAM{PP}{\rm N}
+\hat \HAM{PP}{\rm 1PN}
+\hat \HAM{PP}{\rm 2PN}
+ \hat {\rm H}_{\rm SO}^{\rm LO}
+ \hat {\rm H}_{\rm SO}^{\rm NLO}
+ \hat {\rm H}_{\rm S^2}^{\rm LO}
+ \hat {\rm H}_{\rm S_1 S_2}^{\rm LO}
\,.
\end{eqnarray}
These are sufficient for maximally rotating black holes up to and including 2.5\pN.
The variables
$\unsc{p}_a$ and $\unsc{x}_a$ are the linear canonical momentum and
position vectors,  respectively. They commute with the spin vectors
${\unsc S}_a$,
where ``$a$'' denotes the particle label, $a=1,2$.
$\HAM{\rm PP}{}$ is the conservative point-particle  ADM Hamiltonian known up to 3\pN , see,
e.g., \cite{Jaranowski:Schafer:1998} and \cite{Damour:Jaranowski:Schafer:2001}. 
The LO spin dependent contributions are well-known, see, e.g., \cite{Barker:OConnell:1975,Barker:OConnell:1979,Poisson:1998}.
$\HAM{\rm SO}{\rm NLO}$ 
was recently found in \cite{Damour:Jaranowski:Schafer:2008:1,Steinhoff:Schafer:Hergt:2008} and
$\HAM{\rm S_1 S_2}{NLO}$
 in \cite{Steinhoff:Schafer:Hergt:2008, Steinhoff:Hergt:Schafer:2008:2}
(the latter was confirmed in \cite{Levi:2008}).
The leading-order ${\rm S}_1^2$ and ${\rm S}_2^2$
Hamiltonians were
derived in \cite{DEath:1975} and \cite{Thorne:Hartle:1985}.
%
Measuring the GW signal, determinantion of constraints of the equation of state of both extended bodies is possible in principle.
Hamiltonians of cubic and higher order in spin are given in \cite{Hergt:Schafer:2008:2,Hergt:Schafer:2008},
and higher \pN\  orders linear in spin are tackled in \cite{Steinhoff:Schafer:2009:2,Steinhoff:Wang:2009}.

The four-dimensional model behind the Hamiltonians linear in the single spin variables
is given by the Mathisson-Papapetrou equations \cite{Papapetrou:1951, Mathisson:1937}
\begin{eqnarray}
	\frac{{\rm D} S_a^{\mu\nu}}{{\rm d} \tau} &= 2 p^{[\mu}_a u^{\nu]}_a \,, \\
	\frac{{\rm D} p_a^{\mu}}{{\rm d} \tau} &= - \frac{1}{2} {\rm R}^{\mu}_{~\rho\beta\alpha}
		u^{\rho}_a S^{\beta\alpha}_a \,,
\end{eqnarray}
and the Tulczyjew stress-energy tensor density \cite{Tulczyjew:1959,Dixon:1979}
\begin{equation}
\sqrt{-g} T^{\mu\nu} = \sum_a \int \rmd \tau \bigg[
	u^{(\mu}_a p^{\nu)}_a \delta_{(4)a}
	+ ( u^{(\mu}_a S^{\nu)\alpha}_a \delta_{(4)a} )_{;\alpha}
\bigg] \,,
\end{equation}
which can be used as the source of the gravitational field in the Einstein equations
(see \cite{Steinhoff:Puetzfeld:2009} and references therein for spin-squared corrections in the stress-energy tensor).
Here, the four-dimensional coordinate of the $a$-th object is denoted by $x_a^{\mu}$ and
$p_{a\mu}$ is the linear momentum, $u_{a}^{\mu}$ the 4-velocity, normalised as $ u^\mu u_\mu =-1$, $\tau$ the
proper time parameter, $S^{\mu\nu}_a$ the spin tensor,
``$;$''
denotes the four-dimensional\ covariant derivative,
and $\delta_{(4)a} = \delta(x - x_a)$
with normalisation $\int \rmd^4x \, \delta_{(4)a} = 1$.
${\rm R}^{\mu}_{~\rho\beta\alpha}$ is the four-dimensional\ Riemann
tensor and ${\rm D}/{\rm d}\tau$ the absolute derivative, which is a derivative in direction of the 4-velocity of the (massive) particle.
In order to close the system of equations, one has to impose a spin supplementary condition (SSC), 
which is most conveniently taken to be
\begin{equation}
S^{\mu\nu}_a p_{a\nu} = 0 \,.
\end{equation}
To linear order in spin, $p_{a\mu}=m_a u_{a\mu}$, where $m_a$ is the mass parameter of the $a$-th object.
Notice that the matter variables appearing in the Mathisson-Papapetrou equations
and the stress-energy tensor are related to the canonical variables appearing in the
Hamiltonians by rather complicated redefinitions.

We are going to work in the centre-of-mass (COM) frame, where the total linear momentum
vector is zero, i.e.
$\unsc{p}_2=-\unsc{p}_1=-\unsc{\Pone}$. The Hamiltonians taken into account
depend on 
$\unsc{x}_1$ and $\unsc{x}_2$ only in the combinations
$\unsc{x}_1 - \unsc{x}_2$,
so they can be re-expressed in terms of
$\vct{n}_{12} = -\vct{n}_{21}=\hat{\vct{x}}/\hat{\rm \rel} = \vct{x}/\rel$, the normalised direction from particle 1 to 2,
and $\hat{\rm \rel}=|\unsc{x}_1 - \unsc{x}_2|$ with $\unsc{x}=\unsc{x}_1 - \unsc{x}_2$.

We will make use of the following scalings to convert quantities with hat to dimensionless ones,
 \begin{eqnarray}
  \rm {H}&~\equiv ~&\frac{\hat{\rm H}}{\mu c^2}  					\,,\label{eq::Hscale} \\
 \vct{x}	&~\equiv~&   \unsc{x} \, \left(\frac{G\,m}{c^2} \right)^{-1}	\hrulefill	\,,\label{eq::xscale} \\
 \vct{p}	& ~\equiv~&	 \unsc{p} \, \left( \mu \, c  \right)^{-1}		\,,\label{eq::pscale} \\
 \vct{S}_a & ~\equiv~&  \unsc{S}_a \, \left( \frac{G\, m_a}{c^2} \, (m_a\, c)  \right)^{-1}	\,.\label{eq::Sscale} 
 \end{eqnarray}
Here, $m\equiv m_1+m_2$ denotes the total mass  and
$\mu \equiv m_1\,m_2/m$
is  the reduced mass. The speed of light is denoted by $c$ and $G$ is Newton's gravitational constant.
Additionally, we introduce the reduced orbital angular momentum vector $\vAng \equiv \rel \vnunit \times \vct{p}$ and its norm $\ang \equiv |\vAng|$.

 Explicitly, the contributions to the rescaled version of 
\Eref{eq:dimH} read
%
\begin{eqnarray}
\label{H_full}
\fl \HAM{PP}{N}
&=& \frac{\vPone^2}{2} - \frac{1}{\rel}
\,,
\\
\fl
\HAM{PP}{1PN}
&=& \epso^2 \left\{
\frac{1}{8} (3\eta-1) \left( \vPone^2 \right)^2
- \frac{1}{2} \left[ (3+\eta) ({\vPone}^2) + \eta \scp{\nunit}{\vPone}^2 \right] \frac{1}{\rel}
+ \frac{1}{2\rel^2} \right\}
\,,
\\
\fl
\HAM{PP}{2PN}
&=& {\epso^4} 
\Biggl\{
\frac{1}{16} \left( 1 - 5\eta + 5 \eta^2 \right)
\left( {\vPone}^2 \right)^3
+ \frac{1}{8} \biggl[
\left( 5 - 20 \eta - 3 \eta^2 \right)
\left({\vPone}^2 \right)^2
\neanl
\fl
&&\quad\quad
- 2 \eta^2 \scp{\nunit}{\Pone}^2 \left({\vPone}^2\right)
 - 3 \eta^2
\scp{\nunit}{\Pone}^4 \biggr] \frac{1}{\rel}
+ \frac{1}{2} 
\left[ 
(5 + 8 \eta)\left({\vPone}^2\right) + 3 \eta \scp{\nunit}{\Pone}^2 
\right] \frac{1}{\rel^2}
\nonumber
\\
\fl
&&\quad
- \frac{1}{4} ( 1 + 3 \eta) \frac{1}{\rel^3}
\Biggr\}
\,,
\\
\fl \HAM{SO}{LO}
&=&
\epso^2 \,
\spino
\frac{\aso}{\rel^3} \left\{ \left(1 - \frac{\eta}{2} + \sqrt{1 - 4 \eta}\right) \scp{\ang}{\Sone} +\
\left(1 - \frac{\eta}{2} - \sqrt{1 - 4 \eta}\right) \scp{\ang}{\Stwo} \right\} 
\,,
\\
\fl
\HAM{SO}{NLO}
&=&
\epso^4
\,
\spino
\,
\frac	{\aso}
	{16 \rel^4}
\biggl\{
 \scp{\ang}{\Sone} \biggl[12 \eta \, \rel \left(1-\eta +\sqrt{1-4 \eta }\right)  \scp{\nunit}{\Pone}^2
\neanl
\fl
&& \quad
+\eta \, \rel \left(9 -6 \eta +19
   \sqrt{1-4 \eta }\right)  \left(\vPone^2\right)
\neanl
\fl
&&\quad
-16 \left((\eta +3) \sqrt{1-4 \eta }+3\right)\biggr]
\nonumber \\
\fl
&& - \scp{\ang}{\Stwo} \biggl[12 \eta \, \rel \left(-1 + \eta +\sqrt{1-4 \eta }\right)  \scp{\nunit}{\Pone}^2
\neanl
\fl
&&\quad+\eta \, \rel \left(-9 + 6 \eta +19
   \sqrt{1-4 \eta }\right)  \left(\vPone^2\right)
\neanl
\fl
&&\quad-16 \left((\eta +3) \sqrt{1-4 \eta }-3\right) \biggr]
\biggr\}
\,,
\\
\fl
\HAM{S_1 S_2}{LO}
&=&
\epso^2
\spino^2
\ass \, \frac	{\eta}
	{ \rel^3}
		\left \{
		 3 \scp{\nunit} {\Sone}
		   \scp{\nunit} {\Stwo}
		 -  \scp{\Sone}  {\Stwo}
		\right \}
\,,
\\
\fl
\HAM{S^2}{LO}
&=&
\epso^2 \,
\spino^2 \,
\frac{ \asq
}{2 \rel^3}
   \biggl\{ \lambda_1 \, \left(-1 + 2 \eta -\sqrt{1-4 \eta }\right)
   \left(
		 3	\scp{\nunit}{\Sone}^2
		-	\scp{\Sone} {\Sone}
   \right)\nonumber\\
\fl
&& \quad\quad + \lambda_2 \,  \left(
		-1 + 2 \eta +\sqrt{1-4\eta }\right)
		 \left(3 \scp{\nunit}{\Stwo}^2
			-\scp{\Stwo}{\Stwo} \right) \biggr\}
\,,
\end{eqnarray}
%
where $\eta  \equiv \mu/m$ is the symmetric mass ratio. Without loss of generality we assume that $m_1 > m_2$. Such an assumption is necessary, because the spins are scaled with the individual masses in a non-symmetric way. 

We introduced dimensionless  ``book-keeping'' parameters $\epso$ to count the formal $1/c$ order and
$\spino$ to count the spin order (linear or quadratic).
Evaluating all given quantities, those have to be given the numerical value 1.
The parameters
$\aso$, $\ass$, $\asq$ distinguish the spin--orbit, spin(1)--spin(2) and the spin-squared contributions and
can have values $1$ or $0$, depending on whether the reader likes to incorporate the associated interactions.

The spins are denoted by $\vSone$ for object $1$ and $\vStwo$ for object $2$.
Notice that the S$_1^2$ and S$_2^2$ Hamiltonians depend on constants $\lambda_1$ and
$\lambda_2$, respectively, which parametrise the quadrupole deformation of the objects 1 and 2
due to the spin
and take different values for, e.g., black holes and neutron stars. For black holes, $\lambda_a = - \frac{1}{2}$ and
for neutron stars, $\lambda_a$ can take continuous values from the interval $[-2,-4]$  \cite{Poisson:1998, Laarakkers:Poisson:1999}
\footnote{Note that the definition of the $\lambda_a$ depends
on the definition of the spin Hamiltonian and, thus, can be arbitrarily normalised.
We consistently use the notation mentioned above.}.

The parallelism condition
tells us to set the spins to $\vct{S}_a = \chi_a \vAng/\ang$, where $-1 < \chi_a < 1$. 
During our calculations, we insert the condition of maximal rotation ($S_a \sim \epso$) to cut off every quantity after 2.5\pN,
but list our results in formal orders $S_a \sim \epso^0$ (for the formal counting, see, e.g.,  \cite{Hergt:Schafer:2008} and also Appendix A of \cite{Steinhoff:Wang:2009}).
However, for $S_a \sim \epso^2$, many spin contributions are of the order ${\cal O}(\epso^6)$, i.e. 3\pN\, which
is beyond our present 2\pN\ PP dynamics.
The reader may insert either $S_a \sim \epso$ (maximal rotation) or $S_a \sim \epso^2$ (slow rotation).
{
}
The next step is to evaluate the EOM due to these Hamiltonians and to
find a parametric solution. As stated, we will restrict ourselves to parallel
or anti-parallel
angular momenta and will, finally, only have to take care of the motion
in the orbital plane.


\section{Conservation of parallelism of \texorpdfstring{$\vAng$}{\ang} and \texorpdfstring{$\vSone$}{S1} and \texorpdfstring{$\vStwo$}{S2}}
\label{sec::ConservationOfParallel}
The motion of binaries with arbitrarily-oriented spins is, in general,
chaotic as soon as the spin-spin interaction is included \cite{Levin:2000, Sohr:2009}.
For special configurations, despite this, it is possible to integrate
the EOM analytically, which particularly is the case for aligned spins and orbital angular momentum.

The time derivatives of the spins $\vct{S}_a$ and the total angular momentum $\vct{J}$ are governed by the Poisson brackets with the total Hamiltonian, given by
\begin{eqnarray}
\label{eq::PBS1}
\fl \pb{\vct{\Sone}}{\HAM{}{}} & = & \spino ^2 \epso ^2 \biggl\{\ass\frac{ \eta}{\rel^3} \left(3 \scp{\Stwo}{\nunit}  \crp{\nunit}{\Sone}+\crp{\Sone}{\Stwo}\right)\neanl
\fl	& &\quad+\asq \frac{ \crp{\nunit}{\Sone}}{\rel^3}\scp{\Sone}{\nunit}3 \lambda_1 \left(2 \eta - 1 - \sqrt{1-4 \eta } \right)\biggr\}\neanl
\fl	& &+\spino  \biggl\{\aso \epso ^2 \frac{\crp{\ang}{\Sone}}{\rel^3} \left(-\frac{\eta }{2}+\sqrt{1-4 \eta }+1\right) \neanl
\fl	& &\quad+\aso \epso ^4 \biggl[\frac{\crp{\ang}{\Sone}}{\rel^3} \biggl(\frac{3 }{4} \scp{\Pone}{\nunit}^2 \eta \left(1 -\eta+ \sqrt{1-4 \eta } \right)\neanl
\fl	& &\quad\quad+\frac{1}{16} (\vPone^2) \eta \left(9 -6\eta+19 \sqrt{1-4 \eta }\right)\biggr) \neanl
\fl	& &\quad\quad-\frac{\crp{\ang}{\Sone}}{\rel^4}\left(3 + (\eta+3)\sqrt{1-4 \eta }\right) \biggr]\biggr\}
\,,  \\
 \label{eq::PBS2}
\fl  \pb{\vct{\Stwo}}{\HAM{}{}}&=& \pb{\vct{\Sone}}{\HAM{}{}} (1 \leftrightarrow 2)\,,  \\
\label{eq::PBJ}
\fl \pb{\vct{J}}{\HAM{}{}} & = & \pb{\vct{\ang}}{\HAM{}{}} + \pb{\vct{\Sone}}{\HAM{}{}} + \pb{\vct{\Stwo}}{\HAM{}{}} = 0\,.
\end{eqnarray}
Furthermore, the magnitudes of the spins are conserved, because the spins commute with the linear momentum and the position vector and fulfill the canonical angular momentum algebra.
Note  that the operation $(1 \leftrightarrow 2)$ switches the label indices
of the individual particles and goes along with $\vnunit \leftrightarrow \vct{n}_{21} = -\vnunit$. 
\Eref{eq::PBJ} is not displayed completely here like \Eref{eq::PBS1}.
If we assume parallel spins and orbital angular momentum at $t=0$, all the above Poisson brackets
vanish exactly.
Anyway, this is insufficient to conclude the conservation of parallelism of $\vAng$ and the spins for all times
$t>0$ since
\begin{eqnarray}
\vSone(t)
	&=& \vSone |_{t=0}  + \pb{\vSone}{\HAM{}{}} |_{t=0}\, t + \frac{1}{2} \pb{\pb{\vSone}{\HAM{}{}}}{\HAM{}{}} |_{t=0} \, t^2 + ... \neanl
	&=& \sum_{n=0}^{\infty} \frac{1}{n!} \pb{\vSone}{\HAM{}{}}_{n} |_{t=0} \, t^n\,,
\end{eqnarray}
where
\begin{eqnarray}
\pb{\vSone}{\HAM{}{}}_{n} &=& \pb{\pb{\vSone}{\HAM{}{}}_{n-1}}{\HAM{}{}}\,, \neanl
\pb{\vSone}{\HAM{}{}}_{0} &=& \vSone\label{eq:multipoissonbrackets}\,.
\end{eqnarray}
Because the system of variables $\vSone$ and $\vStwo$ has to be completed with $\vct{\rel}$ and $\vPone$ to characterise
the matter system, one has to give clear information about the full system of EOM.
It is important that, even with vanishing Poisson
brackets of $H$ with $\vSone$ and $\vStwo$, $\vct{\rel}$ and $\vPone$ do change due to the orbital revolution. Thus, one has
to clarify if this non-stationary subsystem of the EOM is able to cause violation of  the parallelism condition during time evolution.
From the stability theory of autonomous ordinary differential equations it is well known that there is a fixed point if {\em all}
time derivatives of the system vanish.
In the case of a system starting at such a fixed point at $t=0$ it will not be able to evolve away from this point.
The discussion of these issues is the main point in the following two subsections.

%

\subsection{Discussion via conservation of constraints}\label{subsec:conservationconstraints}
One way to show the non-violation of the initial constraint of $\vAng \parallel \vSone, \vStwo$ due to the motion of the binary is to argue via the time derivatives of the constraints. These should be written as a linear combination of the constraints themselves. Let
\begin{eqnarray}
\label{Eq::constrainteqns} C_a (x, p, S) &=& 0\,, 
\end{eqnarray}
be the initial constraints of the system. Dirac \cite[p. 36]{Dirac:1964} argued: If one can write
\begin{eqnarray}
\label{Eq::constraintlinearcomb} \dot{C}_a &=& \sum_{b} D_{ab} (x, p, S) C_b\,,
\end{eqnarray}
for the time derivatives of the constraints, the constraints are conserved. That is due to the fact
that every time derivative of \Eref{Eq::constraintlinearcomb} generates only new time derivatives
of the constraints on the one hand, which can be expressed as a linear combination of constraints,  or time
derivatives of the quantities appearing in $D_{ab}$ times the constraints on the other.

In our case the constraints read
\begin{eqnarray}
\label{soneconstr}\vSone - \frac{|\vSone|}{\ang} \vAng = &  \vSone - \tilde{\chi}_1 \vAng &=0\,,\eanl
\label{stwoconstr}\vStwo  - \frac{|\vStwo|}{\ang} \vAng  = &   \vStwo - \tilde{\chi}_2 \vAng &=0	\,,
\end{eqnarray}
with $\tilde{\chi}_1$ and $\tilde{\chi}_2$ denoting the ratios of the spin lengths and the orbital angular momentum.
In general, the quantities $\tilde{\chi}_a$ have non-vanishing time derivatives,
\begin{eqnarray}
\diff{\tilde{\chi}_a}{t} &=& -\frac{|\vct{S}_a|}{\ang^3} \scp{\dot{\ang}}{\ang} = -\tilde{\chi}_a  \frac{\scp{\dot{\ang}}{\ang}}{\ang^2} \,.
\end{eqnarray}
Due to the conservation of the total angular momentum $\vct{J}$, the derivatives of \eref{soneconstr} and \eref{stwoconstr} can be expressed via
\footnote{The reader should be aware that, for a general discussion, spins and orbital angular
momentum have to be scaled with the same quantity to preserve the conservation of the
total angular momentum. In our case, the scaling is different,
but the discussion is only of structural nature and the mass coefficients emerging from a differing
scaling can be absorbed into the appearing factors.}
\begin{eqnarray}
\label{Eq::D_Constr_S1_Dt}
\fl \diff{}{t}\left(\vct{S}_a - \tilde{\chi}_a \vAng\right) &=& \diff{\vct{S}_a}{t} + \tilde{\chi}_a  \frac{\scp{\dot{\ang}}{\ang}}{\ang^2} \vAng - \tilde{\chi}_a \diff{\vAng}{t}\neanl
&=& \diff{\vct{S}_a}{t} - \tilde{\chi}_a \left(\mathbf{1} - \frac{\vAng}{\ang} \otimes \frac{\vAng}{\ang}\right) \diff{\vAng}{t}\neanl
&=&  \diff{\vct{S}_a}{t} + \tilde{\chi}_a \left(\mathbf{1} - \frac{\vAng}{\ang} \otimes \frac{\vAng}{\ang}\right) \sum_b \diff{\vct{S}_b}{t}
\,,
\end{eqnarray}
where the tensor product $1/\ang^2 \vAng \otimes \vAng$ is the projector onto the $\vAng$ direction. 
Note that the constraint equations only depend on the spin derivatives in linear manner.
Hence, it is sufficient to analyse the structure of \Eref{eq::PBS1},
\begin{eqnarray}
\label{Eq::DS1Dt_simple}
\fl \diff{\vSone}{t} & = & D_1 \scp{\Stwo}{\nunit} \crp{\nunit}{\Sone} + D_2 \crp{\Sone}{\Stwo}\neanl
		&& + D_3 \scp{\Sone}{\nunit} \crp{\nunit}{\Sone} + D_4 \crp{\ang}{\Sone}
\,.
\end{eqnarray}
The coefficients $D_k$  ($k=1,\dots,4$) are all scalar functions of the linear momentum $\vPone$,
the separation $\rel$ and other intrinsic quantities.
We are allowed to add vanishing terms to \Eref{Eq::DS1Dt_simple}, namely
\begin{eqnarray}
\fl \diff{\vSone}{t} & = & D_1 \left[\scp{\Stwo}{\nunit} - \tilde{\chi}_2 \scp{\ang}{\nunit}\right] \crp{\nunit}{\Sone} + D_2 \crp{\Sone}{\Stwo}\neanl
		&& + D_3 \left[\scp{\Sone}{\nunit} - \tilde{\chi}_1 \scp{\ang}{\nunit} \right] \crp{\nunit}{\Sone} + D_4 \crp{\ang}{\Sone}
\,.
\end{eqnarray}
As well, we can add a term to the $D_2$ coefficient and subtract it at the end, getting
\begin{eqnarray}
\fl \diff{\vSone}{t} & = & D_1 \left[\scp{\Stwo}{\nunit} - \tilde{\chi}_2 \scp{\ang}{\nunit}\right] \crp{\nunit}{\Sone}\neanl
		&& + D_2 \left[\crp{\Sone}{\Stwo} - \tilde{\chi}_2 \crp{\Sone}{\ang}\right] \neanl
		&& + D_3 \left[\scp{\Sone}{\nunit} - \tilde{\chi}_1 \scp{\ang}{\nunit} \right] \crp{\nunit}{\Sone} \neanl
		&& + (D_4 - D_2 \tilde{\chi}_2) \crp{\ang}{\Sone}
\,.
\end{eqnarray}
Finally, we can insert a vanishing term into the modified last one:
\begin{eqnarray}
\label{Eq::DSoneDt}
\fl \diff{\vSone}{t} & = & D_1 \left[\scp{\Stwo}{\nunit} - \tilde{\chi}_2 \scp{\ang}{\nunit}\right] \crp{\nunit}{\Sone}\neanl
		&& + D_2 \left[\crp{\Sone}{\Stwo} - \tilde{\chi}_2 \crp{\Sone}{\ang}\right] \neanl
		&& + D_3 \left[\scp{\Sone}{\nunit} - \tilde{\chi}_1 \scp{\ang}{\nunit} \right] \crp{\nunit}{\Sone} \neanl
		&& + (D_4 - D_2 \tilde{\chi}_2) \left[\crp{\ang}{\Sone} - \tilde{\chi}_1 \crp{\ang}{\ang}\right]
\,.
\end{eqnarray}
We still need to compute the time derivative of $\vStwo$ to obtain the full derivative of the constraints.
Therefore, let $E_k$ be the scalar coefficients in $\diffl{\vStwo}{t}$ (equivalent to the $D_k$ in \eref{Eq::DSoneDt}). Using this, we can
rewrite  it as
\begin{eqnarray}
\fl \diff{\vStwo}{t} & = & E_1 \left[\scp{\Sone}{\nunit} - \tilde{\chi}_1 \scp{\ang}{\nunit}\right] \crp{\nunit}{\Stwo}\neanl
		&& + E_2 \left[\crp{\Sone}{\Stwo} - \tilde{\chi}_1 \crp{\ang}{\Stwo}\right] \neanl
		&& + E_3 \left[\scp{\Stwo}{\nunit} - \tilde{\chi}_2 \scp{\ang}{\nunit} \right] \crp{\nunit}{\Stwo} \neanl
		&& + (E_4 + E_2 \tilde{\chi}_1) \left[\crp{\ang}{\Stwo} - \tilde{\chi}_2 \crp{\ang}{\ang}\right]
\,.
\end{eqnarray}
Thus, the complete time derivative of e.g. the $\vSone$ constraint \eref{Eq::D_Constr_S1_Dt}  is given by
\begin{eqnarray}
\fl \diff{}{t}\left(\vSone - \tilde{\chi}_1 \vAng\right) &=& \left((1 + \tilde{\chi}_1)\mathbf{1} - \tilde{\chi}_1 \frac{\vAng}{\ang} \otimes \frac{\vAng}{\ang}\right) \neanl
&&\quad\cdot\biggl\{D_1 \left[\scp{\Stwo}{\nunit} - \tilde{\chi}_2 \scp{\ang}{\nunit}\right] \crp{\nunit}{\Sone}\neanl
&&\quad\quad + D_2 \left[\crp{\Sone}{\Stwo} - \tilde{\chi}_2 \crp{\Sone}{\ang}\right] \neanl
&&\quad\quad + D_3 \left[\scp{\Sone}{\nunit} - \tilde{\chi}_1 \scp{\ang}{\nunit} \right] \crp{\nunit}{\Sone} \neanl
&&\quad\quad + (D_4 - D_2 \tilde{\chi}_2) \left[\crp{\ang}{\Sone} - \tilde{\chi}_1 \crp{\ang}{\ang}\right]\biggr\}\neanl
&&+ \tilde{\chi}_1 \left(\mathbf{1} - \frac{\vAng}{\ang} \otimes \frac{\vAng}{\ang}\right) \neanl
&&\quad\cdot\biggl\{E_1 \left[\scp{\Sone}{\nunit} - \tilde{\chi}_1 \scp{\ang}{\nunit}\right] \crp{\nunit}{\Stwo}\neanl
&&\quad\quad + E_2 \left[\crp{\Sone}{\Stwo} - \tilde{\chi}_1 \crp{\ang}{\Stwo}\right] \neanl
&&\quad\quad + E_3 \left[\scp{\Stwo}{\nunit} - \tilde{\chi}_2 \scp{\ang}{\nunit} \right] \crp{\nunit}{\Stwo} \neanl
&&\quad\quad + (E_4 + E_2 \tilde{\chi}_1) \left[\crp{\ang}{\Stwo} - \tilde{\chi}_2 \crp{\ang}{\ang}\right]\biggr\}\,.
\end{eqnarray}
In each of the summands of $E_k$ and $D_k$ in the above equation, one can factor out the constraints {linearly}.
Thus, they do vanish if the constraints are inserted.

\subsection{Discussion via symmetry arguments}\label{subsec:symmetryarguments}
To underline the results of subsection \ref{subsec:conservationconstraints} 
we want to show that the multi Poisson brackets \eref{eq:multipoissonbrackets} all vanish if we demand the
parallelism of $\vAng$ and the spins, as a complement to the constraint evolution analysis.
During the calculation of the expressions, 
we truncated the terms to quadratic order in spin and to 2.5\pN\ order, counting the spin maximally rotating.
There is a finite set of terms which are axial vectors and linear or quadratic in spin. Additionally, the spin has to appear in a vector product.
The reason is that  the spins commute with the PP Hamiltonians, and Poisson brackets of spins with spin Hamiltonians will give cross products of spins
with angular momentum $\vAng$ or $\vSone$, $\vStwo$, respectively. In the Hamiltonians, there are only scalar products of spins with
other vectors or with the spins themselves, so that the $\epsilon_{ijk}$ are still remaining after evaluation of the Poisson brackets.
The products with the correct symmetry and linear and quadratic in spin
 are of the form
\begin{eqnarray*}
\underbrace{\vct{A}}_{\begin{array}{c}
\vct{A}\times\vct{A}\\ \vct{P}\times\vct{P}\end{array}}\cdot\underbrace{S}_{\begin{array}{c}\vct{A}\cdot\vct{A}\\\vct{P}\cdot\vct{P}\end{array}}& \rm{and} & \underbrace{\vct{P}}_{\vct{A}\times\vct{P}}\cdot \underbrace{PS}_{\vct{A}\cdot\vct{P}} 
\end{eqnarray*}
where $\vct{A}$ stands for axial vectors, $\vct{P}$ for polar vectors, $S$ for scalars and $PS$ for pseudo
scalars. The vector $\vct{A}$ can be $ \vAng , \vSone , \vStwo $ which are axial vectors and $\vct{P}$ can
be $ \vnunit ,  \vPone$ which are polar vectors. 
Now we enumerate all spin products that emerge when we compute the first multi Poisson brackets,
omitting scalar factors like e.g. functions of $r$ and $\eta$:
\begin{eqnarray}
\fl \pb{\vSone}{\HAM{}{}} &: & \crp{\Sone}{\nunit} \scp{\Sone}{\nunit}; \crp{\Sone}{\nunit} \scp{\Stwo}{\nunit};\neanl
		& &\crp{\Sone}{\Stwo}; \crp{\ang}{\Sone}\,, \eanl
\fl \pb{\pb{\vSone}{\HAM{}{}}}{\HAM{}{}} &: & \crp{\Sone}{\nunit} \scp{\Pone}{\Sone};\crp{\Sone}{\nunit} \scp{\Pone}{\Stwo};\neanl
			& & \crp{\Sone}{\Pone} \scp{\nunit}{\Sone};\crp{\Sone}{\Pone} \scp{\nunit}{\Stwo}\,, \eanl
\fl \pb{\vSone}{\HAM{}{}}_{3} &: & \crp{\Sone}{\Pone} \scp{\Pone}{\Sone}; \crp{\Sone}{\Pone} \scp{\Pone}{\Stwo};\neanl
\fl		& & \crp{\Sone}{\Pone} \scp{\nunit}{\Sone}; \crp{\Sone}{\Pone} \scp{\nunit}{\Stwo}
\,.
\
\end{eqnarray}
Additionally, terms coming from Poisson brackets of $\vStwo$ with $\HAM{}{}$ appear and can be computed from above with
the operation ($1 \leftrightarrow 2$).
Evaluation of higher multi Poisson brackets will not create new terms, but only increase the number of already known factors in the products.
These terms vanish identically if we consider the parallelism between $\vAng$ and the spins because of
$\scp{\ang}{\nunit} = 0$ and $\scp{\ang}{\Pone} = 0$, by construction.
Due to the vanishing Poisson bracket of $\vct{J} = \vAng + \vSone + \vStwo$ with the Hamiltonian $\HAM{}{}$
(this is true even without demanding the parallelism), the disappearance of all multi Poisson brackets of
$\vSone$ and $\vStwo$ with $\HAM{}{}$ turns out to be sufficient to conclude the constancy
of $\vAng$ if the motion starts with $\vAng \parallel \vSone, \vStwo$.

\section{Kepler Parameterisation}
\label{sec::KeplerParam}

In Newtonian dynamics the Keplerian parameterisation of a compact binary
is a well-known tool for celestial mechanics, see e.g. \cite{Goldstein:1981}.
After going to spherical coordinates in the COM,
($ r, \theta, \phi$) with the associated orthonormal vectors ($\ver, \vetheta,
\vephi$) and restricting to the $\theta=\pi/2$ plane, the Keplerian
parameterisation has the following form:
\begin{eqnarray}
r		&=& a \, (1 - e \, \cos u)	\,,							 \\
\phi - \phi_0	&=& v			\,,							 \\
v		&=& 2\, \arctan \left [
					\sqrt{ \frac{1+e}{1-e} } \, \tan \frac{u}{2}
				\right ]		\,.
\end{eqnarray}
Here, $a$ is the semimajor axis,  $e$ is the numerical eccentricity,  $u$
and $v$ are eccentric and true anomaly, respectively. The time dependency
of $r$ and $\phi$ is given by the Kepler equation,
\begin{equation}
\mean = n \, (t-t_0) = u - e \, \sin u \,,
\end{equation}
where $\mean$ is the mean anomaly and $n$ the so-called mean motion, defined
as  $n\equiv\frac{2\, \pi}{P}$ with  $P$ as the orbital period \cite{Colwell:1993}. In these
formulae $t_0$ and $\phi_0$ are some initial instant and the associated
initial phase.
In terms of the conserved quantities $E$, which is the scaled energy (see \Eref{eq::Hscale})
and numerically identical to $\HAM{}{}$, and the orbital angular momentum $\ang$, the
orbital elements $e$, $a$ and $n$ satisfy
\begin{eqnarray}
a&=&  \frac{1}{2|E|}\,,	\\
e^2&=& 1 - 2\ang^2 |E|\,,			\\
n&=& (2|E|)^{3/2}\,.
\end{eqnarray}
For higher \pN\  accurate EOM it is possible to get a solution in a
perturbative way, having the inverse speed of light as the perturbation
parameter.
The 1\pN\  accurate {\em Keplerian like} (from now on we refer to {\em quasi-Keplerian}) parameterisation was first found in \cite{Damour:Deruelle:1985}
and extended for non-spinning compact binaries in \cite{Schafer:Wex:1993, Memmesheimer:Gopakumar:Schafer:2004} 
to 2\pN\ and finally 3\pN\ accuracy.

In the recent past a number of efforts has been undertaken  to obtain a solution to the problem of spinning compact binaries via calculating the EOM for spin-related angular variables in harmonic gauge.
	For circular orbits, including radiation reaction (RR), the authors of \cite{Mikoczi:Vasuth:Gergely:2005} evaluated several
	contributions to the frequency evolution and the number of accumulated GW cycles up to 2\pN,
	{such as from the spin}, mass quadrupole and the magnetic dipole moment parts.
The gravitational wave form
amplitudes as functions of separations and velocities up to and including 1.5\pN\ PP and 1.5\pN\ SO
corrections are given in \cite{Majar:Vasuth:2006}, discussed for the extreme mass ratio limit in the Lense-Thirring approximation
and later in \cite{Vasuth:Majar:2007} and \cite{Majar:Vasuth:2008} for comparable mass binaries.
Recently, in \cite{Gergely:2009} a set of independent variables and their EOM, characterising the angular momenta, has been provided.

For circular orbits with arbitrary spin orientations and leading-order spin-orbit interactions, 
the spin and orbital solutions for slightly differing masses were given in \cite{Tessmer:2009}.
Including LO contributions of ${\rm S}^2$, ${\rm S}_1 {\rm S}_2$ and SO as well as the Newtonian and 1\pN\ contributions to
the EOM, a certain time-averaged orbital parameterisation was found in \cite{Keresztes:Mikoczi:Gergely:2005}, for
a time scale where the spin orientations are almost constant, but arbitrary and the radial motion has been determined.
Symbolically, those solutions suggest the following form for the quasi-Keplerian
 parametrisation including spin interactions:
%
\begin{eqnarray}
\fl r &=& a_r \, (1-e_r \cos u)\,,						\\
\fl  n (t-t_0) &=& u - e_t \sin u	+ \mc{F}_{v-u} (v-u)
			+ \mc{F}_{v} \sin   v 
			+ \mc{F}_{2v} \sin 2v 
			+ \mc{F}_{3v} \sin 3v +
\dots \,,	\\
\fl \frac{2 \pi}{\Phi} \,(\phi-\phi_0)
   &=& v 		+ \mc{G}_{2v} \sin 2v
			+ \mc{G}_{3v} \sin 3v
			+ \mc{G}_{4v} \sin 4v
			+ \mc{G}_{5v} \sin 5v + \dots
 \,,\\
\fl v &=& 2 \, \arctan	\left[ \sqrt{ \frac{1+e_\phi}{1-e_\phi} }
			    \, \tan \frac{u}{2}
			\right]
\,.
\label{eq:trueanomaly}
\end{eqnarray}
The coefficients $\mc{F}_{\dots}$, $\mc{G}_{\dots}$ are \pN\  functions
of $E$, $\ang$ and $\eta$.
At the end of the calculation for binary dynamics with spin, they will obviously include
spin dependencies as well.

In case RR is included, the orbital elements are not longer to be regarded
as constants. Damour, Gopakumar and Iyer published equations of motion for these elements for
the case that the RR is a small effect and the time derivatives due to RR will not contribute
to the GW expressions explicitly \cite{Damour:Gopakumar:Iyer:2004, Konigsdorffer:Gopakumar:2006}.
Spin effects of RR in eccentric orbits were discussed in \cite{Gergely:Perjes:Vasuth:1998}
and references therein.

\section{The quasi-Keplerian parameterisation for aligned spinning compact binaries}
\label{sec::QKPSpin}
Having proven constancy  in time of the directions of angular momenta, we can adopt
the choice of spherical coordinates 
with $\vAng \parallel \vetheta$ (in the $\theta = \pi/2$ plane) and 
the basis ($\vct{\nunit} = \ver, \vephi$).
Hamilton's equations of motion dictate
\begin{eqnarray}
\label{Eq::EOM_standard_r}
          \dot \rel &=\vct{\nunit} \cdot \dot{\vct{\rel}}=\vct{\nunit} \cdot \frac{\partial \, \HAM{}{}}{\partial \, \vct {\Pone} } \,, \\
\label{Eq::EOM_standard_phi}
\rel \, \dot \phi &=\vephi \cdot \dot{\vct{\rel}}=\vct{\vephi} \cdot \frac{\partial \, \HAM{}{}}{\partial \, \vct {\Pone} } \,,
\end{eqnarray}
%
with $\dot \rel = \diffl{r}{t}$ and $\dot \phi = \diffl{\phi}{t}$, as usual. The next standard
step is to introduce $s\equiv 1/\rel$, such that ${\dot \rel=-\dot s/s^2}$.
Using Equations \eref{Eq::EOM_standard_r} and  \eref{Eq::EOM_standard_phi},
we obtain a relation for ${\dot \rel}^2$ and thus ${\dot s}^2$ and another one for
${\dot \phi/\dot s=\diffl{\phi}{s}}$, where the polynomial of ${\dot s}^2$ is of third
degree in $s$. To obtain a formal 2\pN\ accurate parameterisation\footnote{
When we talk about a formal solution at 2\pN\ here, we mean that we incorporate all terms
up to the order $\epso^{4}$ where the spins are formally counted of order $\epso^0$.}, we first concentrate
on the radial part and search for the two nonzero roots of ${\dot s}^2=0$,
namely $\sp$ and $\sm$. The results, to Newtonian order,  are
\begin{eqnarray}
\label{eq::sp}	\sp & = & \frac{1}{a_r (1 - e_r)} = \frac{1 + \sqrt{1 - 2 \ang^2 |E|}}{\ang^2} + {\cal O}(\epso^2)\,, \eanl
\label{eq::sm}	\sm & = & \frac{1}{a_r (1 + e_r)} = \frac{1 - \sqrt{1 - 2 \ang^2 |E|}}{\ang^2} + {\cal O}(\epso^2)\,,
\end{eqnarray}
$\sm$ representing periastron and $\sp$ as the apastron. Next, we factorise ${\dot s}^2$
with these roots and obtain the following two integrals for the elapsed time $t$
and the total radial period $P$,
\begin{eqnarray}
\label{Eq::RadialPeriod}
	P &=& 2 \, \int_{\sm}^{\sp} \frac{{\cal P}_5(\tau) \rm{d}\tau}{\tau^2 \sqrt{(\tau - \sm)(\sp - \tau)}}\,,
\end{eqnarray}
which is a linear combination of integrals of the type
\begin{eqnarray}
	I^{\prime}_n &= 2\int_{\sm}^{\sp} \frac{\tau^n \rm{d}\tau}{\tau^2 \sqrt{(\tau - \sm)(\sp - \tau)}}\,.
\end{eqnarray}
The time elapsed from $s$ to $\sp$,
\begin{eqnarray}
\label{Eq::TimeElapsed}
	 t-t_0	&=& \, \int_{s}^{\sp} \frac{{\cal P}_5(\tau) \rm{d}\tau}{\tau^2 \sqrt{(\tau - \sm)(\sp - \tau)}}\,,
\end{eqnarray}
is a linear combination of integrals of the type
\begin{eqnarray}
	I_n &=& \int_{s}^{\sp} \frac{\tau^n \rm{d}\tau}{\tau^2 \sqrt{(\tau - \sm)(\sp - \tau)}}\,.
\end{eqnarray}
Both integrals $I_n$ and $I^{\prime}_n$ are given in \ref{AppA} in terms of $\sp$ and $\sm$ for $I^{\prime}$ and in terms of $a_r$, $e_r$, $u$ and $\tilde{v}$ for $I$, respectively.
The function ${\cal P}_5 (s)$ is a fifth order polynomial in $s$ and the factor
$2$ follows from
the fact that from $\sm$ to $\sp$ it is only a half revolution.
With the help of the quasi-Keplerian parameterisation
\begin{equation}
\rel = a_r \, (1 - e_r \, \cos u) \,,
\end{equation}
where $a_r$ and $e_r$ are some 2\pN\ accurate semi-major axis and radial eccentricity,
respectively, satisfying
\begin{eqnarray}
a_r &=& \frac{1}{2} \, \frac{\sp + \sm}{\sm\,\sp} \,,\eanl
e_r &=& \frac{1}{2} \, \frac{\sp - \sm}{\sm +\sp} \,,
\end{eqnarray}
due to \eref{eq::sp} and \eref{eq::sm},
we obtain a 2\pN\ accurate expression for $a_r$ and $e_r$ in terms of several intrinsic quantities.
With \Eref{Eq::TimeElapsed}, we get a preliminary expression for the Kepler Equation, as
we express $n\,(t-t_0) = \frac{2\,\pi}{P}(t-t_0)$ in terms of $u$, and as standard, we introduce
an auxiliary variable 
\begin{equation}
\label{Eq::TildeVDef}
 {\tilde v} \equiv 2 \arctan \left[ \sqrt{\frac{1+e_r}{1-e_r}} \tan \frac{u}{2}\right] \,.
\end{equation}
At this stage, we have
\begin{eqnarray}
\label{Eq::ProtoKE}
 \mean &\equiv& n\,(t-t_0) \neanl
   &=     & u + {\tilde{\cal F}}_{u} \, \sin u + {\tilde{\cal F}}_{\tilde v - u} (\tilde v - u) + {\tilde{\cal F}}_{\tilde v}\, \sin \tilde v\,,
\end{eqnarray}
with $\tilde{\cal F}_{\dots}$ as some 2\pN\ accurate functions of $E$, $\ang$, $\eta$, $\lambda_a$ and $\chi_a$.
These functions are lengthy and only temporarily needed in the derivation of later results, so we will not provide them.

Let us now move on to the angular part. As for the time variable, we factorise the polynomial of
$\diffl{\phi}{s}$ with the two roots $\sm$ and $\sp$ and obtain the elapsed phase at $s$
and the total phase $\Phi$ from $\sm$ to $\sp$,
\begin{eqnarray}
\label{Eq::PhiElapsed}
\phi-\phi_0	= \int_{s}^{\sp}	\frac{{\cal B}_3 ({\tau}) }
						{\sqrt{(\sm -\tau)(\tau -\sp)}}	{\rm d} \tau \,,	\\
\label{Eq::PhiTotal}
\Phi= 2 \int_{\sm}^{\sp}	\frac{{\cal B}_3 ({\tau}) }
						{\sqrt{(\sm -\tau)(\tau -\sp)}}	{\rm d} \tau \,,
\end{eqnarray}
where the function ${\cal B}_3 ({\tau})$ is a polynomial of third order in $\tau$, respectively.
Using \Eref{Eq::PhiElapsed} and \ref{Eq::PhiTotal}, the elapsed phase scaled by the total phase $\frac{2\,\pi}{\Phi}\,(\phi-\phi_0)$ in terms of $\tilde v$
is computed as
\begin{eqnarray}
\label{Eq::ProtoAngular}
\frac{2 \pi}{\Phi} (\phi-\phi_0)
&= \tilde v	+ {\tilde{\cal G}}_{\tilde v} \sin   \tilde v
  		+ {\tilde{\cal G}}_{2\tilde v} \sin 2 \tilde v
		+ {\tilde{\cal G}}_{3\tilde v} \sin 3 \tilde v \,.
\end{eqnarray}
For the following, we change from the auxiliary variable $\tilde{v}$ to the true anomaly
due to \Eref{eq:trueanomaly} with
\begin{equation}
\label{Eq::VExpand}
e_\phi = e_r (1 + \epso^2 c_1 + \epso^4 c_2) \,,
\end{equation}
differing from the radial eccentricity by some 1\pN\ and 2\pN\ level corrections $c_1$ and $c_2$. 
These corrections are fixed in such a way that the $\sin v$ contribution
in $\frac{2 \pi}{\Phi} (\phi-\phi_0) $ vanishes at
each \pN{} order, resulting in a formal lowest order correction with respect to $v$
at 2\pN.
Therefore, we eliminate $u$ in \Eref{Eq::TildeVDef} with the help of  \eref{eq:trueanomaly}
and insert the result into \eref{Eq::ProtoAngular}. 
For convenience of the reader, we give a 2PN accurate expansion
of the expression for $\tilde{v}$ in terms of $v$:
\begin{eqnarray}
\label{Eq::vtildesubst}  \tilde{v}&=&v+\epso ^2 c_1 \frac{  e_r }{e_r^2-1} \sin v\neanl
&&+\epso ^4 \left\{\left(c_2 -c_1^2 \frac{ e_r^2}{e_r^2-1}\right) \frac{e_r}{e_r^2 - 1} \sin v + \frac{1}{4} c_1^2 \frac{ e_r^2}{\left(e_r^2-1\right)^2} \sin (2 v)\right\}\,.
\end{eqnarray}
Having determined the final expression for $e_\phi$, \Eref{Eq::ProtoAngular} takes the form
\begin{equation}
 \frac{2 \pi}{\Phi} (\phi-\phi_0)=v+ {\cal G}_{2v}\sin 2 v + {\cal G}_{3v}\sin 3 v\,.
\end{equation}
With the help of $v$, we can re-express the preliminary Kepler equation \eref{Eq::ProtoKE} in the form of
\begin{eqnarray}
 \ell &=& n\,(t-t_0)  = u - e_t \sin u + {\cal F}_{v-u} (v - u) + {\cal F}_{v} \sin v \,.
\end{eqnarray}
Here, $e_t$ is the {\em time eccentricity} and simply represents the sum of all terms with the
factor $\sin u$ in $\mean$.
All the orbital quantities will be detailed in the next section.

\section{Summarising the results}
\label{sec::Summary}
We present all the orbital elements  $a_r, e_r, e_t, e_\phi, n$ and the functions ${\cal F}_{\dots}$ and ${\cal G}_{\dots}$ of the 
quasi-Keplerian parameterisation
 \begin{eqnarray}
	\rel&=& a_r (1- e_r \cos u) \,, \label{eq::radialparm}\eanl
	n \left(t-t_0\right)  &=& u - e_t \sin u + {\cal F}_{v-u} (v-u)  + {\cal F}_{v} \sin v \,, \label {eq::timeparm}\eanl
	\frac{2 \pi}{\Phi }  \left(\phi -\phi _0\right)&=&v + {\cal G}_{2 v} \sin (2 v) + {\cal G}_{3 v} \sin (3 v)\,, \label{eq::angleparm} \eanl
 v &=& 2\, \arctan \left [ {\sqrt{\frac{1+e_{\phi}}{1-e_{\phi}}} \, \tan \frac{u}{2}} \right ] \,, \label{eq::vparm}
\end{eqnarray}
in the following list.
For $\spino = 0$ (remember that $\spino$ counts the spin order) one recovers the results from, e.g.  \cite{Schafer:Wex:1993}.

\begin{eqnarray}
\fl a_r&=&\frac{1}{2 |E|}\neanl
	\fl&&+\epso ^2 \biggl\{\frac{\eta -7}{4}+\frac{\spino}{\ang} \aso \left[ \sqrt{1-4 \eta } \left(\chi _1-\chi _2\right)+ \left(1-\frac{\eta }{2}\right) \left(\chi _1+\chi _2\right)\right]\neanl
	\fl&&\quad+\frac{\spino ^2}{\ang^2} \biggl[\left(\chi _1-\chi _2\right){}^2 \left(\frac{\ass \eta }{4}+\asq \left(\frac{1}{8} \sqrt{1-4 \eta } \left(\lambda _1-\lambda _2\right)+\frac{1}{8} (1-2 \eta ) \left(\lambda _1+\lambda _2\right)\right)\right)\neanl
	\fl&&\quad\quad+\left(\chi _1+\chi _2\right){}^2 \left(\asq \left(\frac{1}{8} \sqrt{1-4 \eta } \left(\lambda _1-\lambda _2\right)+\frac{1}{8} (1-2 \eta ) \left(\lambda _1+\lambda _2\right)\right)-\frac{\ass \eta }{4}\right)\neanl
	\fl&&\quad\quad+\asq \left(\chi _1+\chi _2\right) \left(\chi _1-\chi _2\right) \left(\frac{1}{4} (1-2 \eta ) \left(\lambda _1-\lambda _2\right)+\frac{1}{4} \sqrt{1-4 \eta } \left(\lambda _1+\lambda _2\right)\right)\biggr]\biggr\}\neanl
	\fl&&+\epso ^4 \biggl\{|E| \biggl(\frac{1}{8} \left(\eta ^2+10 \eta +1\right)\neanl
	\fl&&\quad\quad+\frac{\spino}{\ang} \aso \left[\frac{1}{8}  \left(-6 \eta ^2+19 \eta -8\right) \left(\chi _1+\chi _2\right)+\frac{1}{8}  \sqrt{1-4 \eta } (5 \eta -8) \left(\chi _1-\chi _2\right)\right]\biggr)\neanl
	\fl&&\quad+ \frac{\spino}{\ang^3} \aso \left[\left(\eta ^2-\frac{39 \eta }{4}+8\right) \left(\chi _1+\chi _2\right)+\frac{1}{4} (32-9 \eta ) \sqrt{1-4 \eta } \left(\chi _1-\chi _2\right)\right]\neanl
	\fl&&\quad+\frac{1}{4\ang^2}  (11 \eta -17)\biggr\}\,, \eanl
\fl e_r^2&=&1-2 \ang^2 |E|\neanl
	\fl&&+\epso ^2 \biggl\{\ang^2 |E|^2 \biggl(5 (\eta -3)+\frac{\spino}{\ang} \aso \left[8  \sqrt{1-4 \eta } \left(\chi _1-\chi _2\right)+ (8-4 \eta ) \left(\chi _1+\chi _2\right)\right]\neanl
	\fl&&\quad\quad+\frac{\spino^2}{\ang^2} \biggl[\left(\chi _1-\chi _2\right){}^2 \left(2 \ass \eta  +\asq \left(\sqrt{1-4 \eta } \left(\lambda _1-\lambda _2\right) -(2 \eta -1) \left(\lambda _1+\lambda _2\right) \right)\right)\neanl
	\fl&&\quad\quad\quad+\left(\chi _1+\chi _2\right){}^2 \left(\asq \left(\sqrt{1-4 \eta } \left(\lambda _1-\lambda _2\right) -(2 \eta -1) \left(\lambda _1+\lambda _2\right) \right)-2 \ass \eta  \right)\neanl
	\fl&&\quad\quad\quad+\asq \left(\chi _1+\chi _2\right) \left(\chi _1-\chi _2\right) \left(2 \sqrt{1-4 \eta } \left(\lambda _1+\lambda _2\right) -2 (2 \eta -1) \left(\lambda _1-\lambda _2\right) \right)\biggr]\biggr)\neanl
	\fl&&\quad+|E| \biggl(-2 (\eta -6)+\frac{\spino}{\ang} \aso \left[4  (\eta -2) \left(\chi _1+\chi _2\right)-8 \sqrt{1-4 \eta } \left(\chi _1-\chi _2\right)\right]\neanl
	\fl&&\quad\quad-\frac{\spino^2}{\ang^2} \biggl[\left(\chi _1-\chi _2\right){}^2 \left(2 \ass \eta  +\asq \left(\sqrt{1-4 \eta } \left(\lambda _1-\lambda _2\right) -(2 \eta -1) \left(\lambda _1+\lambda _2\right) \right)\right)\neanl
	\fl&&\quad\quad\quad+\left(\chi _1+\chi _2\right){}^2 \left(\asq \left(\sqrt{1-4 \eta } \left(\lambda _1-\lambda _2\right) -(2 \eta -1) \left(\lambda _1+\lambda _2\right) \right)-2 \ass \eta  \right)\neanl
	\fl&&\quad\quad\quad+\asq \left(\chi _1+\chi _2\right) \left(\chi _1-\chi _2\right) \left(2 \sqrt{1-4 \eta } \left(\lambda _1+\lambda _2\right) -2 (2 \eta -1) \left(\lambda _1-\lambda _2\right) \right)\biggr]\biggr)\biggr\}\neanl
	\fl&&+\epso ^4 \biggl\{\ang^2 |E|^3 \biggl(-4 \eta ^2+55 \eta -80\neanl
	\fl&&\quad\quad+\frac{\spino}{\ang} \aso \left[ \left(6 \eta ^2-49 \eta +80\right) \left(\chi _1+\chi _2\right)+(80-19 \eta ) \sqrt{1-4 \eta } \left(\chi _1-\chi _2\right)\right]\biggr)\neanl
	\fl&&\quad+\frac{|E|}{\ang^2} \biggl(-22 \eta +34\neanl
	\fl&&\quad\quad+\frac{\spino}{\ang} \aso \left[ \left(-8 \eta ^2+78 \eta -64\right) \left(\chi _1+\chi _2\right)+2  \sqrt{1-4 \eta } (9 \eta -32) \left(\chi _1-\chi _2\right)\right]\biggr)\neanl
	\fl&&\quad+|E|^2 \biggl(\eta ^2+\eta +26\neanl
	\fl&&\quad\quad+\frac{\spino}{\ang} \aso \left[ \left(10 \eta ^2-70 \eta +4\right) \left(\chi _1+\chi _2\right)+4  (1-4 \eta )^{3/2} \left(\chi _1-\chi _2\right)\right]\biggr)\biggr\}\,, \eanl
\fl n & = & 2 \sqrt{2} |E|^{3/2}+ \epso ^2 \frac{|E|^{5/2} (\eta -15) }{\sqrt{2}}\neanl
	\fl&& + \epso ^4 \biggl\{4 \frac{|E|^3}{\ang} \biggl(6 \eta -15\neanl
	\fl&&\quad\quad+\aso \frac{\spino}{\ang}  \left[2 \left(\eta ^2-8 \eta +6\right) \left(\chi _1+\chi _2\right)-4 \sqrt{1-4 \eta } (\eta -3) \left(\chi _1-\chi _2\right)\right]\biggr)\neanl
	\fl&&\quad+\frac{|E|^{7/2}}{8 \sqrt{2}}\left(11 \eta ^2+30 \eta +555\right) \biggr\}\,,\label{Eq::n} \eanl
\fl e_t^2 & = &1-2 \ang^2 |E|+\epso ^2 \biggl\{|E| \left(\ang^2 |E| (17-7 \eta )+ 4 (\eta -1) \right)\neanl
	\fl& &\quad+\frac{\spino}{\ang} \aso |E| \left[2 (\eta -2) \left(\chi _1+\chi _2\right) -4  \sqrt{1-4 \eta } \left(\chi _1-\chi _2\right) \right]\neanl
	\fl& &\quad+\frac{\spino ^2 |E|}{\ang^2} \biggl[\left(\chi _1-\chi _2\right){}^2 \left(\asq \left(\left(\eta -\frac{1}{2}\right) \left(\lambda _1+\lambda _2\right) -\frac{1}{2} \sqrt{1-4 \eta } \left(\lambda _1-\lambda _2\right) \right)-\ass \eta  \right)\neanl
	\fl& &\quad\quad+\left(\chi _1+\chi _2\right){}^2 \left(\ass \eta  +\asq \left(\left(\eta -\frac{1}{2}\right) \left(\lambda _1+\lambda _2\right) -\frac{1}{2} \sqrt{1-4 \eta } \left(\lambda _1-\lambda _2\right) \right)\right)\neanl
	\fl& &\quad\quad+\asq \left(\chi _1+\chi _2\right) \left(\chi _1-\chi _2\right) \left((2 \eta -1) \left(\lambda _1-\lambda _2\right) -\sqrt{1-4 \eta } \left(\lambda _1+\lambda _2\right) \right)\biggr]\biggr\}\neanl
	\fl& &+\epso ^4 \biggl\{\frac{|E|}{\ang^2} \biggl(-11 \eta +17+  \ang^4 |E|^2 \left(-16 \eta ^2+47 \eta -112\right) +12 \sqrt{2}  \ang^3 |E|^{3/2} (5-2 \eta )\neanl
	\fl& &\quad\quad +2  \ang^2 |E| \left(5 \eta ^2+\eta +2\right) +6 \sqrt{2} \ang \sqrt{|E|} (2 \eta -5) \biggr)\neanl
	\fl& &\quad+\frac{\spino}{\ang} \aso \frac{|E|}{2 \ang^2} \biggl[ \left(\chi _1+\chi _2\right)  \biggl(-16 \sqrt{2}  \ang^3 |E|^{3/2} \left(\eta ^2-8 \eta +6\right) +\ang^2 |E| \left(32 \eta ^2-159 \eta +124\right) \neanl
	\fl& &\quad\quad\quad+8 \sqrt{2}\sqrt{|E|} \ang \left(\eta ^2-8 \eta +6\right) -8 \eta ^2+78 \eta -64\biggr)\neanl
	\fl& &\quad\quad+ \sqrt{1-4 \eta } \left(\chi _1-\chi _2\right) \biggl(32 \sqrt{2}  \ang^3 |E|^{3/2} (\eta -3) + \ang^2 |E| (124-59 \eta ) \neanl
	\fl& &\quad\quad\quad-16 \sqrt{2}  \ang \sqrt{|E|} (\eta -3) +18 \eta -64\biggr)\biggr]\biggr\} \,, \eanl
\fl \mathcal{F}_{v-u} & = & -\epso ^4 \frac{2 \sqrt{2} |E|^{3/2}}{\ang} \biggl\{ 3  \left( \eta -\frac{5}{2}\right) \neanl
	\fl&&\quad+\frac{\spino}{\ang} \aso \left[  \left(\eta ^2-8 \eta +6\right)  \left(\chi _1+\chi _2\right)- 2 \sqrt{1-4 \eta } (\eta -3)  \left(\chi _1-\chi _2\right)\right]\biggr\}\,, \eanl
\fl \mathcal{F}_{v} & = & \epso ^4 \frac{|E|^{3/2}}{2 \sqrt{2} \ang} \sqrt{1-2 \ang^2 |E|} \biggl\{-\eta  (\eta +4) \neanl
	\fl&&\quad-\frac{\spino}{\ang} \aso  \biggl[\sqrt{1-4 \eta } (\eta +8) \left(\chi _1-\chi _2\right) - (13 \eta -8)  \left(\chi _1+\chi _2\right)\biggr]\biggr\}\,, \eanl
\fl  \frac{\Phi }{2 \pi }&=&1+\epso ^2 \frac{1}{\ang^2} \biggl\{3+\frac{\spino}{\ang} \aso \left[ (\eta -2) \left(\chi _1+\chi _2\right)-2 \sqrt{1-4 \eta } \left(\chi _1-\chi _2\right)\right] \neanl
	\fl &&	+\frac{\spino ^2}{\ang^2} \biggl[\frac{3}{8} \left(\chi _1-\chi _2\right){}^2 \left(-2 \ass \eta -\asq \sqrt{1-4 \eta } \left(\lambda _1-\lambda _2\right)+\asq (2 \eta -1) \left(\lambda _1+\lambda _2\right)\right) \neanl
	\fl &&	\quad+\frac{3}{8} \left(\chi _1+\chi _2\right){}^2 \left(2 \ass \eta -\asq \sqrt{1-4 \eta } \left(\lambda _1-\lambda _2\right)+\asq (2 \eta -1) \left(\lambda _1+\lambda _2\right)\right) \neanl
	\fl &&	\quad+\frac{3}{4} \asq \left(\chi _1+\chi _2\right) \left(\chi _1-\chi _2\right) \left((2 \eta -1) \left(\lambda _1-\lambda _2\right)-\sqrt{1-4 \eta } \left(\lambda _1+\lambda _2\right)\right)\biggr]	\biggr\} \neanl
	\fl &&	+\epso ^4 \biggl\{\frac{|E|}{\ang^2} \left(3 \eta -\frac{15}{2}+\frac{\spino}{\ang} \aso \left[2  \left(\eta ^2-8 \eta +6\right) \left(\chi _1+\chi _2\right)-4 \sqrt{1-4 \eta } (\eta -3) \left(\chi _1-\chi _2\right)\right]\right) \neanl
	\fl &&	+\frac{1}{\ang^4}\biggl(\frac{\spino}{\ang} \aso \left[\frac{21}{4}  \sqrt{1-4 \eta } (\eta -8) \left(\chi _1-\chi _2\right)-\frac{3}{4} \left(2 \eta ^2-49 \eta +56\right) \left(\chi _1+\chi _2\right)\right] \neanl
	\fl &&	\quad-\frac{15}{4} (2 \eta -7)\biggr)\biggr\}\,,
\eanl
\fl \mathcal{G}_{2 v} & = & \epso ^4 \frac{\left(2 \ang^2 |E|-1\right)}{4 \ang^4} \biggl\{\frac{\eta  (3 \eta -1) }{2}\neanl
	\fl&&\quad+3 \frac{\spino}{\ang} \aso \left[  \sqrt{1-4 \eta } \eta  \left(\chi _1-\chi _2\right) - (\eta -1) \eta  \left(\chi _1+\chi _2\right) \right]\biggr\}\,,
\eanl
\fl \mathcal{G}_{3 v} & = & \epso ^4 \frac{\left(1-2 \ang^2 |E|\right)^{3/2}}{8\ang^4} \biggl\{-\frac{3 \eta ^2 }{4}\neanl
	\fl&&\quad+\frac{\spino}{\ang} \aso \left[(\eta -1) \eta  \left(\chi _1+\chi _2\right) - \sqrt{1-4 \eta } \eta  \left(\chi _1-\chi _2\right) \right]\biggr\}\,,
\eanl
\fl e_{\phi }^2 & = &1-2 \ang^2 |E|+\epso ^2 \biggl\{|E| \left(\ang^2 |E|(\eta -15) +12\right)\neanl
\fl	& &\quad+\frac{\spino}{\ang} \aso |E| \left(\ang^2 |E|-1\right) \left[8  \sqrt{1-4 \eta } \left(\chi _1-\chi _2\right) -4  (\eta -2) \left(\chi _1+\chi _2\right) \right]\neanl
\fl	& &\quad+\frac{\spino ^2 |E|}{\ang^2} \left(4 \ang^2 |E|-3\right) \neanl
\fl	& &\quad\quad\times\biggl[\left(\chi _1-\chi _2\right){}^2  \left(\ass \eta  +\asq \left(\frac{1}{2} \sqrt{1-4 \eta } \left(\lambda _1-\lambda _2\right) -\frac{1}{2} (2 \eta -1) \left(\lambda _1+\lambda _2\right) \right)\right)\neanl
\fl	& &\quad\quad+\left(\chi _1+\chi _2\right){}^2  \left(-\ass \eta  +\asq \left(\frac{1}{2} \sqrt{1-4 \eta } \left(\lambda _1-\lambda _2\right) -\frac{1}{2} (2 \eta -1) \left(\lambda _1+\lambda _2\right) \right)\right)\neanl
\fl	& &\quad\quad+\asq \left(\chi _1+\chi _2\right) \left(\chi _1-\chi _2\right) \left(\sqrt{1-4 \eta } \left(\lambda _1+\lambda _2\right) -(2 \eta -1) \left(\lambda _1-\lambda _2\right) \right)\biggr]\biggr\}\neanl
\fl	& &+\epso ^4 \biggl\{\frac{|E|}{8 \ang^2} \left(-4 \ang^4 |E|^2 \left(3 \eta ^2-30 \eta +160\right) +4 \ang^2 |E| \left(9 \eta ^2+88 \eta -16\right) -15 \eta ^2-232 \eta +408\right)\neanl
\fl	& &\quad+\frac{\spino}{\ang}  \aso \frac{ |E|}{2\ang^2} \biggl[\left(\chi _1+\chi _2\right)  \left(2 \ang^4 |E|^2 (80-31 \eta ) +8 \ang^2 |E| \left(\eta ^2-36 \eta +17\right) -3 \left(\eta ^2-71 \eta +64\right)\right)\neanl
\fl	& &\quad\quad+\sqrt{1-4 \eta } \left(\chi _1-\chi _2\right)  \left(-2 \ang^4 |E|^2 (\eta -80) +4 \ang^2 |E| (34-15 \eta ) +33 \eta -192\right)\biggr]\biggr\} 
\,.
 \end{eqnarray}
For the case that one chooses $e_t$ instead of the other eccentricities as the intrinsic parameter to be searched
for in the data analysis investigations, we give the ratios of the other eccentricities with respect to  $e_t$:
\begin{eqnarray}
 \fl \frac{e_r}{e_t} & = & 1+\epso ^2 \biggl\{(8-3 \eta ) |E|\neanl
	\fl&&\quad+\aso \frac{\spino |E|}{\ang} \left[(\eta -2) \left(\chi _1+\chi _2\right) -2 \sqrt{1-4 \eta } \left(\chi _1-\chi _2\right) \right]\neanl
	\fl&&\quad+ \frac{\spino ^2 |E|}{\ang^2} \biggl[\left(\chi _1-\chi _2\right){}^2 \left(\asq \left(\frac{1}{4} (2 \eta -1) \left(\lambda _1+\lambda _2\right) -\frac{1}{4} \sqrt{1-4 \eta } \left(\lambda _1-\lambda _2\right) \right)-\frac{\ass \eta  }{2}\right)\neanl
	\fl&&\quad\quad+\left(\chi _1+\chi _2\right){}^2 \left(\frac{\ass \eta  }{2}+\asq \left(\frac{1}{4} (2 \eta -1) \left(\lambda _1+\lambda _2\right) -\frac{1}{4} \sqrt{1-4 \eta } \left(\lambda _1-\lambda _2\right) \right)\right)\neanl
	\fl&&\quad\quad+\asq \left(\chi _1+\chi _2\right) \left(\chi _1-\chi _2\right) \left(\left(\eta -\frac{1}{2}\right) \left(\lambda _1-\lambda _2\right) -\frac{1}{2} \sqrt{1-4 \eta } \left(\lambda _1+\lambda _2\right) \right)\biggr]\biggr\}\neanl
\fl	&&+\epso ^4 \biggl\{\frac{|E|}{2 \ang^2} \left[\ang^2 |E| \left(6 \eta ^2-63 \eta +56\right) -6 \sqrt{2} \ang \sqrt{|E|} (2 \eta -5) -11 \eta +17\right]\neanl
\fl	&&\quad+ \aso \frac{\spino}{\ang}  \frac{|E|}{4 \ang^2} \Bigl[\sqrt{1-4 \eta } \left(\chi _1-\chi _2\right)  \left(\ang^2 |E| (23 \eta -84) +16 \sqrt{2} \ang \sqrt{|E|} (\eta -3) +18 \eta -64\right)\neanl
\fl	&&\quad\quad-\left(\chi _1+\chi _2\right) \left(\ang^2 |E| \left(8 \eta ^2-55 \eta +84\right) +8 \sqrt{2} \ang \sqrt{|E|} \left(\eta ^2-8 \eta +6\right) +8 \eta ^2-78 \eta +64\right)\Bigr]\biggr\}\,,\eanl
\
\fl \frac{e_{\phi }}{e_t} & = &1+\epso ^2 \biggl\{-2 (\eta -4) |E|\neanl
	\fl&&\quad+\aso \frac{\spino |E|}{\ang} \left[(\eta -2) \left(\chi _1+\chi _2\right) -2 \sqrt{1-4 \eta } \left(\chi _1-\chi _2\right) \right]\neanl
	\fl&&+\frac{\spino ^2 |E|}{\ang^2} \biggl[\left(\chi _1-\chi _2\right){}^2 \left(\asq \left(\left(\eta -\frac{1}{2}\right) \left(\lambda _1+\lambda _2\right) -\frac{1}{2} \sqrt{1-4 \eta } \left(\lambda _1-\lambda _2\right) \right)-\ass \eta  \right)\neanl
	\fl&&\quad\quad+\left(\chi _1+\chi _2\right){}^2 \left(\ass \eta  +\asq \left(\left(\eta -\frac{1}{2}\right) \left(\lambda _1+\lambda _2\right) -\frac{1}{2} \sqrt{1-4 \eta } \left(\lambda _1-\lambda _2\right) \right)\right)\neanl
	\fl&&\quad\quad+\asq \left(\chi _1+\chi _2\right) \left(\chi _1-\chi _2\right) \left((2 \eta -1) \left(\lambda _1-\lambda _2\right) -\sqrt{1-4 \eta } \left(\lambda _1+\lambda _2\right) \right)\biggr]\biggr\}\neanl
\fl	&&+\epso ^4 \biggl\{\frac{|E|}{16 \ang^2} \left[2 \ang^2 |E| \left(11 \eta ^2-168 \eta +224\right) -48 \sqrt{2} \ang \sqrt{|E|} (2 \eta -5) -15 \eta ^2-144 \eta +272\right]
\neanl
\fl&&\quad
+\aso \frac{\spino}{\ang} \frac{|E|}{4 \ang^2} \Bigl[\left(\chi _1+\chi _2\right)  \left(-3 \ang^2 |E| \left(2 \eta ^2-15 \eta +28\right) -8 \sqrt{2} \ang \sqrt{|E|} \left(\eta ^2-8 \eta +6\right) +5 \eta ^2+135 \eta -128\right)\neanl
	\fl&&\quad\quad+\sqrt{1-4 \eta } \left(\chi _1-\chi _2\right) \left(\ang^2 |E| (13 \eta -84) +16 \sqrt{2} \ang \sqrt{|E|} (\eta -3) +15 \eta -128\right)\Bigr]\biggr\}
\,.
\end{eqnarray}

\section{Gravitational wave forms}\label{sec::Gravwave}

The final form of the GW model consists of two ingredients. The first one is the expression
for the far-zone radiation field, which will naturally depend on general kinematic quantities
describing the binary system.
The explicit time evolution of these kinematical quantities represents the second one.

In the first part of this section, we will compute the GW amplitudes $h_\times$
and $h_+$. We will require the associated corrections to the desired order, which are available
in the literature in harmonic coordinates, see below for references.
As well, we will use coordinate transformations from ADM to harmonic coordinates to be
able to apply the time evolution of the orbital elements in the previous sections, which 
we have computed in ADM coordinates only.

In the second part, we will provide the latter quantities as implicit functions of time.


\subsection{\pN\ expansion of the gravitational radiation amplitudes}
The transverse-traceless (TT) projection of the radiation field and thus $h_\times$ and $h_+$, the two
polarisations, strongly
depend on the observer's position relative to the source. To obtain $h_\times$ and $h_+$, it is necessary
to give position relations of the orbital plane to the direction where the detector is situated. 
An observer--dependent coordinate system will be helpful to give the time domain waveform 
expressions in terms of the radial separation, orbital angular velocity and the spins.

\begin{center}
\begin{figure}[ct]
\centering
   \psfrag{X}{\hspace{-.5cm} $\vex = \vct{p}$}
   \psfrag{Y}{$\vey$}
   \psfrag{Z}{$\vez, \vAng/\ang$}
   \psfrag{n12}{$\vnunit$}
   \psfrag{p}{{$\vct{p}$}}
   \psfrag{q}{{$\vct{q}$}}
   \psfrag{N}{\hspace{-0.4cm} {$\vct{N}$}}
   \psfrag{phi2}   {\large $\varphi$}
   \psfrag{phi}    {\large $\phi$}
   \psfrag{i0}     {\hspace{-0.2cm} \vspace{0.9cm} \large $i_0$}
  \psfrag{h}{$\vAng$}
   \psfrag{invariable}	{\bf{\small{invariable plane	}}}
    \psfrag{orbital}  		{\bf{\small{orbital plane}}}
   \psfrag{POS}       		{\bf{\small{\hspace{-0.5cm} plane of sky    	}}}
\psfrag{iS}{i}
  \includegraphics[scale=0.5, angle=-0]{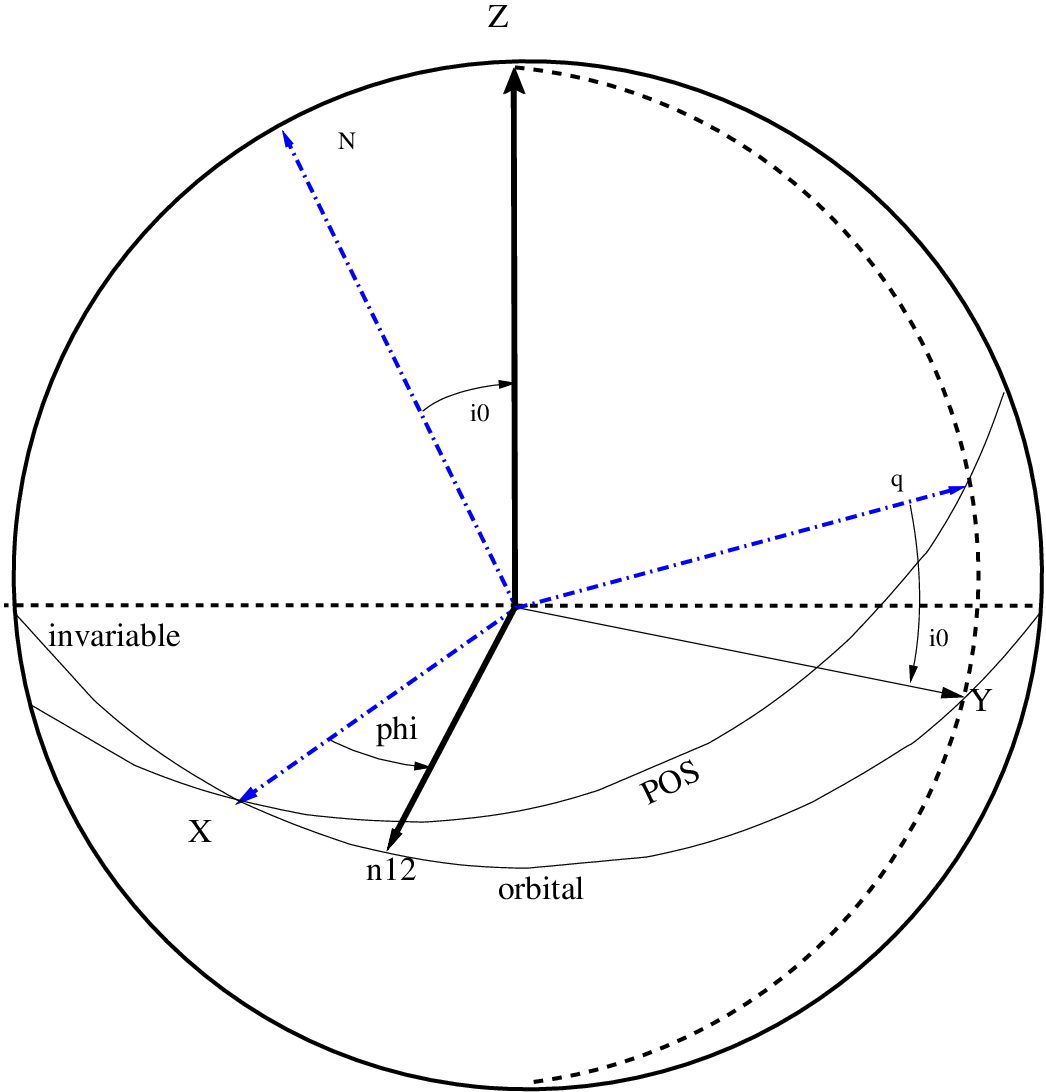}
\caption{
	The geometry of the binary. We have added the observer related
	frame $(\vct{p},\vct{q},\vct{N})$ (in dashed and dotted lines)
	with $\vct{N}$ as the line--of--sight vector. $\vct{N}$ points from
	the origin of the invariable frame $(\vct{e_x}, \vct{e_y}, \vct{e_z} )$
	to the observer. Note that the orbital angular
	momentum $\vAng$ lies on the $\vct{e_z}$ axis and so do the spins.
	$\vct{N}$ is chosen to lie in the $\vct{e_y}$-$\vct{e_z}$--plane,
	and measures a constant angle $i_0$ (associated with the rotation around
	$\vct{e_x}$) from $\vct{e_z}$, such that $\vct{p} = \vct{e_x}$, and this
	is the line where the orbital plane meets the plane of the sky.
	The angle $i_0$ is also found between the vector $\vct{q}$, itself
	positioned in $(\vct{e_y}, \vct{e_z})$, too, and $\vct{e_y}$.
}
\label{fig:figure1}
\end{figure}
\end{center}

We start the calculation by defining the unit line-of-sight-vector $\vct{N}$ as pointing from the source
to the observer. Now, let the unit vectors $\vct{p}$ and $\vct{q}$ span the plane of the sky for the observer
and complete the orthonormal basis ($\vct{p}, \vct{q}, \vct{N}$),
\begin{eqnarray}
\label{Eq::p_times_q_pqN}
\vct{p} \times \vct{q} = \vct{N}\quad{\rm and\;cyclic.}
\end{eqnarray}
Additionally, let us define an invariant reference coordinate system ($\vex, \vey, \vez$).
Both coordinate systems can be coupled by a special orthogonal matrix. We follow \cite{Tessmer:2009}
and construct the triad ($\vct{p}, \vct{q}, \vct{N}$) by a rotation around the vector $\vex$ with
some constant inclination angle $i_0$,
\begin{eqnarray}
\label{Eq::rotation_xyz_pqN}
\left(
\begin{array}{ccc}
\vex\\
\vey\\
\vez
\end{array}
\right)
&=
\left(
\begin{array}{ccc}
1	&	0		&	0	\\
0	&	\cos 	i_0	&	\sin	i_0\\
0	&	-\sin	i_0	&	\cos	i_0
\end{array}
\right)
&
\left(
\begin{array}{c}
\vct{p}\\
\vct{q}\\
\vct{N}
\end{array}
\right)\,.
\end{eqnarray}
\Fref{fig:figure1} shows shows a representation of what has been done. We clearly see that the vector
$\vct{p}$ coincides with $\vex$
	\footnote{
	In reference \cite{Tessmer:2009} the caption for Figure 2 should be made precise.
	The plane of the sky meets the orbital plane at $\vex$ for $\Upsilon=0$ only.
	Generally, at $\vex=\vct{p}$ the plane of the sky meets the invariable plane.
	}.
Next, we express the radial separation $\vct{\rel}$ in the
orbital plane ($\vex, \vey$) and perform the rotation \Eref{Eq::rotation_xyz_pqN}
to move to the observer's triad and calculate $\vct{\rel}$ and $\vct{v}$,
\begin{eqnarray}
\label{Eq::r_in_pqN}
\fl
 \vct{\rel} &=& \rel \, \left(
\vct{p} \, \cos \phi			+
\vct{q} \, \cos i_0 \, \sin \phi	+
\vct{N} \, \sin i_0 \, \sin \phi
\right)\,, \\
\label{Eq::v_in_pqN}
\fl
\vct{v} &= &
	\vct{p} \left( {\dot {\rel}} \cos \phi -{\rel} \dot \phi  \sin \phi 					\right)	
	+\vct{q}  \left({\rel} \dot{\phi}  \cos i_0 \cos \phi +\dot{\rel} \cos i_0 \sin \phi 	\right)	
	+\vct{N} \left({\rel} \dot{\phi}  \sin i_0 \cos \phi +\dot{\rel} \sin i_0 \sin \phi 	\right)\,.
\end{eqnarray}
This provides the orbital contributions to the field.
To compute the spin contributions to the radiation field, we also expand the spins in the orbital triad,
\begin{eqnarray}
\label{Eq::Sone_In_pqN}
 \vSone=&\chi_1 \, \vez =& \chi_1 \, (\vct{N} \cos i_0 - \vct{q} \, \sin i_0) \,, \\
\label{Eq::Stwo_In_pqN}
 \vStwo=&\chi_2 \, \vez =& \chi_2 \, (\vct{N} \cos i_0 - \vct{q} \, \sin i_0) \,.
\end{eqnarray}
We also need to know how $h_\times$ and $h_+$ are extracted from the TT part.
This is done via following projections:
\begin{eqnarray}
 h_\times = \frac{1}{2}\,(p_i q_j  + p_j q_i) \, \ttproj{ij}{kl} \, \htt{kl} \,, \\
 h_+ = \frac{1}{2}\,(p_i p_j  - q_j q_i) \,   \ttproj{ij}{kl} \, \htt{kl} \,,
\end{eqnarray}
where $\ttproj{ij}{kl}$ is the usual TT projector onto the line-of-sight vector $\vct{N}$,
\begin{equation}
 \ttproj{ij}{kl} \equiv 	(\delta_k^i - N^i N_k)
				(\delta_l^j - N^j N_l)
	-\frac{1}{2} (\delta^{i j}-N^i N^j) (\delta_{kl} - N_{k} N_{l})
\,,
\end{equation}
and we define
\begin{eqnarray}
 \pproj{\times}{ij}	\equiv	\frac{1}{2} (q_i p_j+p_i q_j)\label{eq:Cross}	\,,	\eanl
 \pproj{+}{ij}		\equiv	\frac{1}{2} (p_i p_j-q_i q_j)\label{eq:plus}		\,,\label{Eq::px_projectors}
\end{eqnarray}
which are unaffected by the TT projection operator.
The above expressions, \Eref{Eq::p_times_q_pqN}-\eref{Eq::px_projectors}, enable us to compute
all the considered contributions to the radiation field polarisations.
Following  \cite{Kidder:1995} and \cite{Gopakumar:Iyer:1997}, we list the lowest order contributions to the
gravitational waveform in harmonic coordinates. These are the PP contributions to 2\pN, including the NLO-SO
and LO-${\rm S}_1 {\rm S}_2$ terms.
We also add the terms emerging from the gauge transformation from ADM to harmonic coordinates,
\begin{eqnarray}
\label{eq:htt}
\fl
 \htt{ij}&=&\frac{2 \eta}{R'} \, \Biggl[
			\xii{0}{PP}{ij}
+ \epso		\xii{0.5}{PP}{ij}
+ \epso^2	\xii{1}{PP}{ij}
+ \epso^3	\xii{1.5}{PP}{ij}
+ \epso^4	\xii{2}{PP}{ij}
\neanl
\fl&&
\quad \quad
 + \epso^2 \spino \, \aso \,	\xii{1}{SO}{ij}
 + \epso^3 \spino \, \aso \,	\xii{1.5}{SO}{ij}
 + \epso^2 \spino^2 \ass \xii{1}{S_1 S_2}{ij}
\neanl
\fl&&
\quad \quad
+ \epso^2 \spino \,	 \aso \, 
					^{ {\bf g} } \xii{0+1}{PP+SO}{ij}
+ \epso^3 \spino \,	 \aso \, 
					^{ {\bf g} } \xii{0.5+1}{PP+SO}{ij}
+ \epso^4		\,	
					^{ {\bf g} } \xii{0+2}{PP+PP}{ij}
\Biggr]\,.
\end{eqnarray}
Those terms in the last line of the above equation,  labeled ``{\bf g}'', denote corrections coming from
the gauge transformation from ADM to harmonic coordinates to the desired order
\cite{Damour:Jaranowski:Schafer:2008:1, Damour:Jaranowski:Schafer:2001:2}.
\ref{AppB} gives deeper information about how velocities, distances and normal vectors
change within this transformation.
We find it convenient to give a hint to their origin by putting the GW multipole order and the order/type of the
correction in the label, for example ``$(0+1){\rm PP + SO}$'' is the first Taylor correction of the ``Newtonian'' (PP)
quadrupole moment where the coordinates are shifted by a 1\pN\ SO term.

According to \Eref{eq:Cross} and \Eref{eq:plus} one can define the projected components of the $\xi$ via
\begin{eqnarray}
 \xip{{\rm order}}{}{\times} & = & \pproj{\times}{ij} \xii{{\rm order}}{}{ij}\,,\\
 \xip{{\rm order}}{}{+} & = & \pproj{+}{ij} \xii{{\rm order}}{}{ij}\,,
\end{eqnarray}
where the ``cross'' and ``plus'' polarisations read
\begin{eqnarray}
\fl \xip{0}{PP}{\times, + }  & =& 2\biggl\{ \OutXPvv  - \frac{1}{\rel}\, \OutXPnn \biggr \}\,,\label{eq:xipp0}\\
\fl \xip{0.5}{PP}{\times, + } & = & \frac{\delta m}{m}\, \biggl \{ 3 \scp{N}{\nunit} { 1 \over \rel} \left[  
 2\OutXPnv  -\dot \rel \OutXPnn  \right] + 
\scp{N}{v}\left [ { 1 \over \rel} \OutXPnn  
-2 \OutXPvv  \right ] \biggr \}\,, \\ 
\fl \xip{1}{PP}{\times, + } &=&{ 1 \over 3} \biggl \{ (1 -3\eta)
 \biggl [ \scp{N}{\nunit}^2 { 1 \over \rel} 
\biggl ( \left ( 3\vct{v}^2 -15 \dot \rel^2 + 7 { 1 \over \rel} \right )\OutXPnn  + 30 \dot \rel \OutXPnv 
- 14 \OutXPvv  \biggr )
  \nonumber \\ \fl 
& &+ 
\scp{N}{\nunit} \scp{N}{v}{ 1 \over \rel} \left [ 12 \dot \rel \OutXPnn 
 -32 \OutXPnv 
\right ]  
+  \scp{N}{v}^2 \left [ 6 \OutXPvv  -2{ 1\over \rel} \OutXPnn  \right ] \biggr ]
\nonumber \\ \fl
 & & 
+ \biggl [ 3(1 -3\eta) \vct{v}^2 -2(2 -3\eta){ 1 \over \rel} \biggr ] \OutXPvv 
+ 4 { 1 \over \rel} \dot \rel ( 5 +3\eta) \OutXPnv  
\nonumber \\ \fl 
 & &
+ { 1 \over \rel} \biggl [ 3( 1-3\eta)\dot \rel^2 - (10 +3\eta)\vct{v}^2 
+ 29{ 1 \over \rel} \biggr ] \OutXPnn 
\biggr \} \,,\\
\fl \xip{1.5}{PP}{\times, + } &=& \frac{\delta m}{m}\biggl\{\,{ 1 \over 12}\,{(1 -2 \eta)}\biggl \{ 
\scp{N}{\nunit}^3 { 1 \over \rel}
\biggl [   
\left ( 45 \vct{v}^2 -105 {\dot \rel}^2 +90 { 1 \over \rel} \right ) \dot \rel \OutXPnn 
-96 \dot \rel \OutXPvv 
\nonumber \\ \fl
 & &
-\left ( 42 \vct{v}^2 
-210 \dot \rel^2 + 88 { 1 \over \rel} \right ) \OutXPnv  \biggr ]
\nonumber \\
\fl & &
- \scp{N}{\nunit}^2  \scp{N}{v}\,{ 1 \over \rel} \biggl [ \left (27 \vct{v}^2 - 135 \dot \rel^2 
+84 { 1 \over \rel} \right )\OutXPnn 
+ 336 \dot \rel \OutXPnv  -172\OutXPvv  \biggr ]
\nonumber \\
\fl & &
- \scp{N}{\nunit}\scp{N}{v}^2 { 1 \over \rel} \biggl [ 48 \dot \rel \OutXPnn  -184 \OutXPnv  \biggr ]
+ \scp{N}{v}^3 \biggl [ 4 { 1 \over \rel} \OutXPnn  -24 \OutXPvv  \biggr ] \biggl \}
\nonumber \\
\fl & &
- { 1 \over 12}\scp{N}{\nunit}\,{ 1 \over \rel}
\biggl \{ \biggl [ ( 69 -66 \eta) 
\vct{v}^2 - (15 -90\eta)\dot \rel^2 
- ( 242 -24\eta){ 1 \over \rel} \biggr ]\dot \rel \OutXPnn  
\nonumber \\
\fl & &
- \biggl [ ( 66 -36\eta)\vct{v}^2 + (138 + 84\eta)\dot \rel^2 
\nonumber \\
\fl & &
-( 256 - 72\eta){ 1 \over \rel}
\biggr ]\OutXPnv 
+ ( 192 +12 \eta) \dot \rel \OutXPvv  \biggr \}
\nonumber \\
\fl & &
+{ 1 \over 12}\,\scp{N}{v}\biggl \{ \biggl [ ( 23 -10\eta)\vct{v}^2 
-( 9 -18\eta)\dot \rel^2 
-(104 -12\eta){ 1 \over \rel} \biggr ] { 1 \over \rel} \OutXPnn 
\nonumber \\
\fl & &
- \left( 88 +40\eta \right )\,{ 1 \over \rel}\,\dot \rel \OutXPnv 
- \biggl [ ( 12 -60\eta)\vct{v}^2 - ( 20 -52\eta) { 1 \over \rel} \biggr ]\OutXPvv 
\biggr \}\biggr\} \,, \label{eq:1.5pN_hx} \\
\fl \xip{2}{PP}{\times, + } &=& { 1 \over 120}( 1 -5\eta +5 \eta^2) \biggl \{
240\,\scp{N}{v}^4 \OutXPvv  -\scp{N}{\nunit}^4 
\nonumber \\
\fl &&
{ 1\over \rel} \biggl [ \biggl (90 (\vct{v}^2)^2 +
 ( 318 { 1\over \rel} 
-1260 \dot \rel^2)\vct{v}^2 
+ 344 { 1 \over \rel^2} + 1890 \dot \rel^4 
\nonumber \\
\fl & &
-2310 { 1\over \rel} \dot \rel^2 \biggr )\OutXPnn 
\nonumber \\
\fl & &
+ \biggl ( 1620 \vct{v}^2 +3000 { 1\over \rel} - 3780 \dot \rel^2 \biggr ) \dot \rel \OutXPnv 
- \biggl ( 336 \vct{v}^2 - 1680 \dot \rel^2 + 688 { 1\over \rel} \biggr ) \OutXPvv 
\biggr ]
\nonumber \\
\fl & &
-\scp{N}{\nunit}^3 \scp{N}{v} { 1\over \rel} \biggl [ \biggl ( 1440 \vct{v}^2 
- 3360 \dot \rel^2 +
 3600 { 1\over \rel} \biggr )\dot \rel \OutXPnn  
\nonumber \\
\fl & &
-\biggl ( 1608 \vct{v}^2 - 8040 \dot \rel^2 
+ 3864 { 1\over \rel} \biggr ) \OutXPnv  
- 3960 \dot \rel \OutXPvv  \biggr ]
\nonumber \\
\fl & &
+ 120 \scp{N}{v}^3 \scp{N}{\nunit} {1 \over \rel}
 \biggl ( 3 \dot \rel \OutXPnn  -20 \OutXPnv  \biggr )
\nonumber \\
\fl & &
+ \scp{N}{\nunit}^2 \scp{N}{v}^2 { 1\over \rel} \biggl [ \biggl ( 
396 \vct{v}^2 -1980 \dot \rel^2 + 1668 { 1\over \rel} 
\biggr ) \OutXPnn  + 6480 \dot \rel \OutXPnv 
\nonumber \\
\fl & &
-3600 \OutXPvv  \biggr ] \biggr \}
-{ 1 \over 30}\, \scp{N}{v}^2 \biggl \{ 
\biggl [ ( 87 -315 \eta +145 \eta^2 )\vct{v}^2 -( 135 -465\eta + 75 \eta^2 )\dot \rel^2
\nonumber \\
\fl & &
-( 289 -905\eta +115\eta^2 ){ 1\over \rel} \biggr ] {1\over \rel}\,\OutXPnn 
\nonumber \\
\fl & &
- \biggl ( 240 -660\eta -240\eta^2 \biggr )\dot \rel \OutXPnv 
\nonumber \\
\fl & &
-\biggl [ ( 30 -270 \eta +630\eta^2)\vct{v}^2  - 60( 1 - 6\eta 
+10 \eta^2 ){ 1\over \rel}
\biggr ] \OutXPvv \biggr \}
\nonumber \\
\fl & &
+ {1 \over 30} \scp{N}{\nunit} \scp{N}{v} { 1\over \rel} 
\biggl \{ \biggl [ ( 270 
- 1140 \eta +1170 \eta^2 )\vct{v}^2
\nonumber \\
\fl & &
- ( 60 -450 \eta +900 \eta^2 )\dot \rel^2 -( 1270 -3920 \eta 
+360 \eta^2 ){ 1\over \rel} 
\biggr ] \dot \rel \OutXPnn 
\nonumber \\
\fl & &
- \biggl [ ( 186 -810 \eta + 1450\eta^2)\vct{v}^2  +
 (990 -2910\eta -930\eta^2)\dot \rel^2 
\nonumber \\
\fl & &
-( 1242 -4170\eta 
+1930\eta^2){ 1\over \rel}
\biggr ]\OutXPnv 
\nonumber \\
\fl & &
+ \biggl [ 1230 -3810\eta -90\eta^2 \biggr ]\dot \rel \OutXPvv  \biggr \}
\nonumber \\
\fl & &
+ { 1 \over 60} \scp{N}{\nunit}^2 { 1\over \rel} \biggl \{ \biggl [
( 117 -480\eta + 540\eta^2 )(\vct{v}^2)^2 -( 630 -2850\eta +4050\eta^2)\vct{v}^2\dot \rel^2
\nonumber \\
\fl & &
-( 125 -740\eta +900 \eta^2){ 1\over \rel}\,\vct{v}^2 
\nonumber \\
\fl & &
+ ( 105 -1050\eta +3150\eta^2)\dot \rel^4
+( 2715 -8580\eta +1260\eta^2){ 1\over \rel} \,\dot \rel^2 
\nonumber \\
\fl & &
-( 1048 -3120\eta +240\eta^2) { 1\over \rel^2} \biggr ] \OutXPnn 
\nonumber \\
\fl & &
+ \biggl [ ( 216 -1380\eta +4320\eta^2 )\vct{v}^2 
+ ( 1260 -3300\eta -3600 \eta^2) \dot \rel^2 
\nonumber \\
\fl & &
-( 3952 - 12860\eta + 3660\eta^2)
{ 1\over \rel} \biggr ]\,\dot r\, \OutXPnv 
\nonumber \\
\fl & &
- \biggl [ ( 12 -180\eta +1160\eta^2)\vct{v}^2  
+( 1260 -3840\eta -780\eta^2)\dot \rel^2 
\nonumber \\
\fl & &
-( 664 -2360\eta +1700\eta^2){ 1\over \rel} \biggr ] \OutXPvv  \biggr \}
\nonumber \\
\fl & &
- { 1 \over 60}\biggl \{  \biggl [ ( 66 -15 \eta -125\eta^2)(\vct{v}^2)^2
\nonumber \\
\fl & &
+( 90 -180\eta -480\eta^2)\vct{v}^2 \dot \rel^2 -( 389 +1030\eta 
-110\eta^2){ 1\over \rel}\,\vct{v}^2
\nonumber \\
\fl & &
+ ( 45 -225 \eta +225 \eta^2 ) \dot \rel^4 +
( 915 -1440\eta +720\eta^2){ 1\over \rel}\,\dot \rel^2 
\nonumber \\
\fl & &
+ ( 1284 +1090\eta){ 1\over \rel^2} \biggr ]\,{ 1\over \rel}\,\OutXPnn 
\nonumber \\
\fl & &
-\biggl [ ( 132 +540\eta -580\eta^2 )\vct{v}^2 
+( 300 - 1140\eta +300\eta^2)\dot \rel^2
\nonumber \\
\fl & &
+( 856 +400\eta +700\eta^2) { 1\over \rel}
\biggr ]\,{ 1\over \rel }\,\dot r\,\OutXPnv 
\nonumber \\
\fl & &
-\biggl [ ( 45 -315\eta +585\eta^2 )(\vct{v}^2)^2 
+( 354 -210\eta -550\eta^2)\,{ 1\over \rel}\,\vct{v}^2 
\nonumber \\
\fl & &
-( 270 -30\eta +270 \eta^2) { 1\over \rel}\,\dot \rel^2 
\nonumber \\
\fl & &
-( 638 + 1400 \eta -130\eta^2)\,{ 1\over \rel^2}\biggr ] \OutXPvv 
\biggr \}\,,
\end{eqnarray}

\begin{eqnarray}
 \fl \xii{1}{\rm SO}{\times,+}		&=&
-\frac{1}{{\rel}^2} \, \left\{{\SCProjXPnDeltaCrossN}+\sqrt{1-4 \eta }
   {\SCProjXPnSCrossN}\right\}
\,,
\end{eqnarray}

\begin{eqnarray}
 \fl \xii{1.5}{\rm SO}{\times,+}		&=&
 \frac{1}{{ {\rel}^2}}
\,
\Biggl\{
\sqrt{1-4 \eta } \Biggl[6  {\OutXPnn}
\SPAT{\vct{v}}{\vct{\Delta}}{\vnunit}
-6
    {\dot \rel}  {\SCProjXPnDeltaCrossn}
\neanl \fl && 
+4
    {\SCProjXPnDeltaCrossv}
\Biggr]
+6  {\OutXPnn}
\SPAT{\vct{v}}{\vct{S}}{\vnunit}
\neanl \fl &&
+\eta 
    {\SCProjXPnSCrossN} 
\left(6  {\dot \rel}  {\SCNn}-4 {\OutSCNv} \right)
\neanl \fl && 
-6  {\dot \rel}  {\SCProjXPnSCrossn}
\neanl \fl && 
+\eta  \left(4
    {\SCProjXPvSCrossn}
-4  {\SCNn}
{\SCProjXPvSCrossN}
\right)
\neanl \fl &&
+(2 \eta +4)
    {\SCProjXPnSCrossv}
\Biggr\}
\,,
\end{eqnarray}

\begin{eqnarray}
\fl \xip{2}{S_1 S_2}{\times} &=&
 -\frac{3 \eta}{{\rel}^3} \chi _1 \chi _2 \cos \left(i_0\right) \sin (2 \phi )\,,
 \eanl
\fl \xip{2}{S_1 S_2}{+} & = &
- \frac{3 \eta}{4 {\rel}^3} \chi _1 \chi _2 \left(\cos \left(2 i_0\right) \cos (2 \phi )+2 \sin ^2\left(i_0\right)+3 \cos (2 \phi )\right)\,.
\end{eqnarray}

\vspace{1cm}\noindent
The remaining contributions are the gauge dependent terms. Explicitly, they read
%

\begin{eqnarray}
\fl ^{ {\bf g} } \xii{0+1}{PP+SO}{\times,+}		&=&
-\frac{\eta}{ {\rel}^2}
\Biggl\{
3  {\OutXPnn}
\SPAT{\vct{v}}	{\vct{S}}	{\vnunit}
+2
    {\SCProjXPnSCrossv}
+2
    {\SCProjXPvSCrossn}
\Biggr\}
\,,
\end{eqnarray}

\begin{eqnarray}
\fl ^{ {\bf g} } \xii{0.5+1}{PP+SO}{\times,+}		&=&
\frac{{\delta m}}{m}\frac{\eta }{2 \rel^2}
\Biggl\{
 {\OutXPnn} \Bigl[
-15   {\dot \rel}   {\SCNn}   {\SCvSCrossn}-3   {\dot \rel}
     {\SCNSCrossv}
\neanl \fl && 
+3   {\OutSCNv}
     {\SCvSCrossn}-\frac{1}{  {\rel}}  {\SCNSCrossn}
\Bigr]
\neanl \fl && 
+  {\OutSCNn} \Biggl[-6
     {\dot \rel}   {\SCProjXPnSCrossv}+18   {\OutXPnv}   {\SCvSCrossn}
\neanl \fl && 
-\frac{6}{  {\rel}}
     {\SCProjXPnSCrossn}+6   {\SCProjXPvSCrossv}
\Biggr]
\neanl \fl && 
+  {\OutSCNv} \Biggl[2{\SCProjXPnSCrossv}+4   {\SCProjXPvSCrossn} \Biggr]
\neanl \fl && 
+6   {\OutXPnv}   {\SCNSCrossv}+2
     {\OutXPvv}   {\SCNSCrossn}
\Biggr\}
\,,
\end{eqnarray}

\begin{eqnarray}
\label{eq:htt_2PN_gauge}
\fl ^{ {\bf g} } \xii{0 + 2}{PP+PP}{\times,+} &=&
\frac{1}{\rel}
\Biggl\{
{\OutXPnv } \dot{\rel} \left[\frac{1}{2}\eta  \left(3 {\dot \rel}^2-7  {\vct{v}^2}\right)-\frac{2(5 \eta -1) }{{\rel}}\right]
+{\OutXPnn } \left[ \frac{5 \eta  \left({\vct{v}^2}-11{\dot \rel}^2\right)}{4 {\rel}} +\frac{12 \eta +1}{2{ \rel}^2}\right] \neanl \fl
&&
+{\OutXPvv}\left[ \frac{1}{2}\eta  \left(17 {\dot \rel}^2-13 {\vct{v}^2}\right)+\frac{21 \eta +1}{{\rel}}\right]
%
\Biggr\}
\,.
\end{eqnarray}

%
%
\noindent
\Eref{eq:htt_2PN_gauge} shows total agreement with the transformation term in Equation (A2)
of \cite{Damour:Gopakumar:Iyer:2004}.
The next block of equations evaluates the scalar products of vectors and projectors containing the spins.
First, we list those with the total spin $\vct{S}=\vSone+\vStwo$. For those terms with $\vct{\Delta}=\vSone-\vStwo$ instead of $\vct{S}$,
simply replace ($\vct{S} \rightarrow \vct{\Delta}$) on the left hand side and
$(\chi _1+\chi _2) \rightarrow (\chi _1-\chi _2)$ on the right.
The used abbreviations are given by
\begin{eqnarray}
\fl
\SCSnCrossv							&=& {\dot \phi} {\rel} \left(\chi _1+\chi _2\right) 	
\,, \eanl \fl 
\SCNSCrossn							&=& ({\chi_1}+{\chi_2}) \sin ({i_0}) \cos (\phi )																\,, \eanl \fl 
\SPAT{\vct{N}}{\vct{S}}{\vct{v}} & = & \bigl(\chi _1 + \chi _2\bigr) \sin\bigl(i_{0}\bigr) \bigl(\dot{\rel} \cos(\phi ) -  \dot{\phi} \rel \sin(\phi )\bigr) \,, \eanl \fl
 \SCProj{\times}{v^j}{\SCrossn{i}}			&=& \frac{1}{2} \left(\chi _1+\chi _2\right) \cos \left(i_0\right) \left\{{\dot \rel} \cos (2 \phi)-{\dot \phi} {\rel} \sin (2 \phi ) \right\}	 \,,\eanl \fl
  \SCProj{+}{v^j}{\SCrossn{i}}			&=& \frac{1}{8} \left(\chi _1+\chi _2\right) \biggl\{-{\dot \phi} {\rel} \left(\cos \left(2 i_0\right)+3\right) \cos (2 \phi ) +2 {\dot \phi} {\rel} \sin^2\left(i_0\right)\neanl
&&\quad-{\dot \rel} \left(\cos \left(2 i_0\right)+3\right) \sin (2 \phi  )\biggr\}				 \,, \eanl \fl
  \SCProj{\times}{\nunit^i}{\SCrossn{j}}	&=& \frac{1}{2} \left(\chi _1+\chi _2\right) \cos \left(i_0\right) \cos (2 \phi ) 										 \,, \eanl \fl
  \SCProj{+}{\nunit^i}{\SCrossn{j}}		&=& -\frac{1}{8} \left(\chi _1+\chi _2\right) \left\{\cos \left(2 i_0\right)+3\right\} \sin (2 \phi ) 							 \,,\eanl \fl
  \SCProj{\times}{v^i}{\SCrossv{j}}		&=&-\frac{1}{2} \left(\chi _1+\chi _2\right) \cos \left(i_0\right) \left\{2 {\dot \phi}
    {\dot \rel} {\rel} \sin (2 \phi )+\cos (2 \phi ) ({\dot \phi}  {\rel}-{\dot \rel}) ({\dot \phi} {\rel}+{\dot \rel}) \right\}													 \,,\eanl \fl
  \SCProj{+}{v^i}{\SCrossv{j}}			&=& \frac{1}{8} \left(\chi _1+\chi _2\right) \left(\cos \left(2 i_0\right)+3\right) \left\{ \sin (2 \phi ) ({\dot \phi} {\rel}-{\dot \rel}) ({\dot \phi} {\rel}+{\dot \rel})-2
    {\dot \phi} {\dot \rel} {\rel} \cos (2 \phi ) \right\}																										 \,,\eanl \fl
  \SCProj{\times}{\nunit^i}{\SCrossv{j}}	&=& \frac{1}{2} \bigl(\chi _1 + \chi _2\bigr) \cos\bigl(i_{0}\bigr) \bigl\{\dot{\rel} \cos(2 \phi ) -  \
\dot{\phi} \rel \sin(2 \phi )\bigr\} \,, \eanl \fl 
  \SCProj{+}{\nunit^i}{\SCrossv{j}}		&=& \frac{1}{8} \Bigl(\chi _1 + \chi _2\Bigr) \Bigl\{- \dot{\phi} \rel \cos\bigl(2 \phi \bigr) \bigl(3 + \cos(2 i_{0})\bigr) - 2 \dot{\phi} \rel \sin^2(i_{0}) \neanl
&&-  \dot{\rel} \sin\bigl(2 \phi \bigr) \bigl(3 + \cos(2 i_{0})\bigr)\Bigr\} \,, \eanl \fl
  \SCProj{+}{\SCrossN{i}}{\nunit^j} & = & -\frac{1}{2} (\chi _1 + \chi _2)  \sin(i_{0}) \cos(\phi )\,,\eanl \fl
  \SCProj{\times}{\SCrossN{i}}{\nunit^j} & = & -\frac{1}{4} (\chi _1 + \chi _2) \sin(2 i_{0}) \sin(\phi ) \,, \eanl \fl
  \SCProj{+}{\SCrossN{i}}{v^j} & = & -\frac{1}{2} \bigl(\chi _1 + \chi _2\bigr) \sin\bigl(i_{0}\bigr) \bigl(\dot{\rel} \cos(\phi ) - \dot{\phi} \rel \sin(\phi )\bigr) \,,\eanl \fl
  \SCProj{\times}{\SCrossN{i}}{v^j} & = & -\frac{1}{4}\bigl(\chi _1 + \chi _2\bigr) \sin\bigl(2 i_{0}\bigr) \bigl(\dot{\phi} \rel \cos(\phi ) + \dot{\rel} \sin(\phi )\bigr)  \,. 
\end{eqnarray}
The spin-independent projections and the ratio of the difference to the sum of the masses read
\begin{eqnarray} \fl
{\OutSCNn}	&\equiv {{N_i}{\nunit^{i}}}	&=\sin ({i_0}) \sin (\phi ) 																				\,, \eanl \fl
{\OutSCNv}	&\equiv {{N_i}{v^i}}	&=\sin ({i_0}) \left\{{\rel} \dot \phi  \cos (\phi )+\dot{\rel} \sin (\phi ) \right\}												\,, \eanl \fl
 \vct{v}^2			&\equiv {{v^i}{v_i}}	&={\rel}^2 \dot \phi ^2+{\dot \rel}^2																			\,, \eanl \fl
{\OutXnn}	&\equiv \pproj{\times}{ij} {\nunit^i \nunit^j} &=\cos \left(i_0\right) \sin (\phi ) \cos (\phi )														\,, \eanl \fl
{\OutXvv}	&\equiv \pproj{\times}{ij} {v^i v^j}	&=\frac{1}{2} \cos \left(i_0\right) \left\{ \sin (2 \phi ) \left({\dot \rel}^2-{\dot \phi}^2 {\rel}^2\right)+2 {\dot \phi}{\dot \rel} {\rel} \cos (2 \phi )\right\} 
																																		\,, \eanl \fl
{\OutXnv}	&\equiv \pproj{\times}{ij} {\nunit^i v^j}	&=\frac{1}{2} \cos \left(i_0\right) \left\{{\dot \phi} {\rel} \cos (2 \phi )+{\dot \rel} \sin (2 \phi ) \right\}
																																		\,, \eanl \fl
{\OutPnn}	&\equiv \pproj{+}{ij} {\nunit^i \nunit^j}&=\frac{1}{2} \left\{\cos ^2(\phi )-\cos ^2\left(i_0\right) \sin ^2(\phi )\right\}								\,, \eanl \fl
{\OutPvv}	&\equiv \pproj{+}{ij} {v^i v^j}&=\frac{1}{2} \left\{({\dot \rel} \cos (\phi )-{\dot \phi} {\rel} \sin (\phi ))^2-\cos ^2\left(i_0\right) ({\dot \phi} {\rel} \cos (\phi )+{\dot \rel} \sin (\phi ))^2\right\}
																																		\,, \eanl \fl
{\OutPnv}	&\equiv \pproj{+}{ij} {\nunit^i v^j}&=\frac{1}{8} \left\{-{\dot \phi} {\rel} \left(\cos \left(2 i_0\right)+3\right) \sin (2 \phi )-4{\dot \rel} \cos ^2\left(i_0\right) \sin ^2(\phi )+4{\dot \rel} \cos ^2(\phi )\right\} \label{eq:outpnv}																													\,, \eanl \fl
\frac{\delta m}{m} 
			&\equiv  \frac{m_1-m_2}{m} &=\sqrt{1-4 \eta} 																						\,.
\end{eqnarray}
In the expression for the emitted gravitational wave amplitudes, \Eref{eq:htt}, $R^\prime$ is the rescaled distance from the observer to the binary system,
\begin{equation}
 R^\prime=R \, \frac{G m}{c^2}\,.
\end{equation}
We note that it is very important that $R'$ has got the same scaling as $\rel$ in order to remove the physical dimensions.
The common factor $c^{-4}$ of $h^{\rm{TT}}_{ij}$ will be split in $c^{-2}$ for the distance $R'$ and $c^{-2}$
for the $\xii{\dots}{}{}$, in order to make {\em all} terms dimensionless.
Also note that the Equations \eref{eq:xipp0}-\eref{eq:outpnv} in our special coordinates are valid 
only when $\vAng$ is constant in time. 
In the non-aligned case, additional angular velocity contributions kick in and the expressions become rather impractical.
From reference \cite{Arun:Buonanno:Faye:Ochsner:2009}, the reader can extract  explicit higher-order spin
corrections to the Newtonian quadrupolar field for the case of quasi-circular orbits.

\subsection{Dynamical orbital variables as implicit functions of time}

We are now in the position to compute the time domain gravitational wave polarisations with
the help of our orbital elements, to be expressed in terms of conserved quantities and the
mean anomaly, which is an implicit function of time.
Using Equations \eref{eq::radialparm}, \eref{eq::timeparm}, \eref{eq::angleparm} and \eref{eq::vparm}
one can express the quantities $\rel$, $\dot{\rel}$, $\phi$, $\dot{\phi}$ (which are used in the radiation formulas)
in terms of the eccentric anomaly $u$, other orbital elements and several formal 2\pN\ accurate functions. 
The most compact quantity is $\rel$ which is given by \Eref{eq::radialparm}, namely
\begin{eqnarray}
\label{Eq::r_u_short}
\fl \rel(u)& =& a_r (1 - e_r \cos u)  \eanl
\fl &=&
\frac{\dt}{2 |E|} \,
\Biggl[
1
+\epso^2 \, \frac{|E|} {2 \dt} \, \left\{\dt (9-5 \eta )+6 \eta -16\right\}
\neanl \fl &&
 +\epso ^4 \frac{|E|^2}{ \dt} \Biggl \{
 \frac{1}{4 \left(e_t^2-1\right)} \,
 \Bigl[ \dt (\eta  (7 \eta -58)+1) \left(e_t^2-1\right)
\neanl \fl &&
 +2 \eta  \left(-3 \eta  \left(e_t^2-1\right)+34 e_t^2-56\right)+68 \Bigr]
-\frac{6 (\dt-1) (2 \eta -5) }{ \OTS}
\Biggr \}
\neanl \fl && 
 -\epso^2 \, \spino \, \aso \, \frac{|E|^{3/2}}{\dt \OTS }
		 \left\{2 \sqrt{2} \omfeta  \left(\chi _1-\chi _2\right)-\sqrt{2} (\eta -2) \left(\chi _1+\chi _2\right)\right\}
\neanl \fl && 
	+ \frac{\epso^2 \,\spino^2 |E|  ^2}{{\dt}} \times
\neanl \fl &&
\quad 
\Biggl\{
\left(\chi _1-\chi _2\right){}^2
 \left[\frac{\asq \,  \left(
(\lambda_1 + \lambda_2)(2\eta - 1) - (\lambda_1 - \lambda_2) \omfeta
\right)}{2  \left(e_t^2-1\right)}-\frac{\eta  \ass  }{e_t^2-1}\right]
\neanl \fl && 
\quad +\left(\chi _1+\chi _2\right){}^2 \left[\frac{\asq \,  \left(
(\lambda_1 + \lambda_2)(2\eta - 1) - (\lambda_1 - \lambda_2) \omfeta
\right)}{2 \left(e_t^2-1\right)}+\frac{\eta  \ass  }{e_t^2-1}\right]
\neanl \fl && 
\quad +\frac{\asq \,  \left(\chi _1-\chi _2\right) \left(\chi _1+\chi _2\right)} {e_t^2-1}
 \left[
(\lambda_1 - \lambda_2)(2\eta-1) - (\lambda_1 + \lambda_2)\omfeta
\right]
\Biggr\}
\neanl \fl && 
 + \frac{\epso^4 \spino \, |E| ^{5/2}}{2 \sqrt{2}\dt  } \times
\neanl \fl && 
\quad \Biggl \{ \frac{16 (\dt  -1)} {(e_t^2-1)} \, { \left[((\eta -8) \eta +6) \left(\chi _1+\chi _2\right)-2 (\eta -3)\omfeta  \left(\chi _1-\chi
   _2\right)\right]}
\neanl \fl && 
\quad +\frac{4}{\OTS^3} \, \left[3 (\eta  (2 \eta -15)+12) \left(\chi _1+\chi _2\right)-(13 \eta -36)\omfeta  \left(\chi _1-\chi_2\right)\right]
\neanl \fl && 
\quad +\frac{1}{\OTS} \, \left[(19 \eta -42)\omfeta  \left(\chi _1-\chi _2\right)-(\eta  (13 \eta -50)+42) \left(\chi _1+\chi _2\right) \right]
\Biggr \}
\Biggr ]\,.
\end{eqnarray}
Using expression \eref{Eq::r_u_short}, we calculate the derivative via chain rule, given by
\begin{eqnarray}
\label{Eq::dot_r_u_short}
\fl \dot{\rel}(u) &=& \diff{\rel}{u} \diff{u}{t} = n a_r e_r \sin u \times \neanl
\fl && \quad 
\biggl \{
1-{e_t} \cos u  
+ {\cal F}_v \frac{\sqrt{1-e_\phi ^2}  (e_\phi -\cos u  )}{(1-e_\phi  \cos u  )^2}
+{\cal F}_{v-u}\left[ \frac{\sqrt{1-e_\phi ^2}}{1-e_\phi  \cos u  }-1\right]
\biggr \}^{-1}
\eanl
\label{Eq::dot_r_u_long}
\fl
 &=&
\frac{{e_t} \sqrt{2\,|E|} \sin  u}{1-{e_t} \cos u} \,
\Biggl[
	   1
	+ \epso^2 \,  |E|  \, \frac{3}{4} (1-3 \eta )
	+ \epso^4\,{|E|^2}\, \biggl\{
										  \frac{1}{32} (23+\eta  (47 \eta -98))
\neanl
\fl
&&
										+\frac{6 (2 \eta -5) \sqrt{1-e_t^2}}{	\left(e_t^2-1\right) \left(1-e_t \cos  u\right)}
										+\frac{-\frac{1}{2} (\eta -20) \eta -30}{\left(1-e_t \cos  u\right){}^2}
										-\frac{\eta  (\eta +4) \left(e_t^2-1\right)}{2 \left(1-e_t \cos u\right){}^3}
		\biggr\}
\neanl
\fl
&&
+ \aso \, \spino \, \times
\neanl
\fl
&&
\epso^4 \, {|E|}^{5/2} \,
\biggl\{
\frac{4 \sqrt{2}
 \left(
-2 \sqrt{1-4 \eta } (\eta -3) \left(\chi _1-\chi _2\right)
+((\eta -8) \eta +6) \left(\chi _1+\chi _2\right)
\right)}{\left(e_t^2-1\right) \left(1-e_t \cos  u\right)}
\neanl
\fl
&&
\quad \quad \quad
-\frac{
 \left(
\sqrt{1-4 \eta } (17 \eta -40) \left(\chi _1-\chi _2\right)+((51-8 \eta ) \eta -40) \left(\chi _1+\chi_2\right)
\right)}{\sqrt{2-2 e_t^2} \left(1-e_t \cos  u\right){}^2}
\neanl
\fl
&&
\quad \quad \quad
+\frac{
 \sqrt{1-e_t^2} \left(\sqrt{1-4 \eta } (\eta +8) \left(\chi _1-\chi _2\right)+(8-13 \eta ) \left(\chi _1+\chi _2\right)\right)}{\sqrt{2}
   \left(1-e_t \cos  u\right){}^3}
\biggr\}
\Biggr]
\,.
\end{eqnarray}

\noindent
The final expression for $\phi$ in terms of $u$ is rather complicated. 
It is convenient to give a short expression and a description how to
obtain it.
From Equations \eref{eq::angleparm} and \eref{eq::vparm}
one can eliminate $v$ to obtain
\begin{eqnarray}
\label{Eq::phi_u_short}
\fl \phi(u) &=& \phi_0 + \frac{\Phi}{2 \pi}
 \biggl\{
2 \arctan \left[ \sqrt{\frac{1+e_\phi}{1-e_\phi }}
   \tan \frac{u}{2} \right]+{\cal G}_{2 v}  \frac{2 \sqrt{1-e_\phi ^2} \sin (u) (e_\phi -\cos (u))}{(e_\phi  \cos
   (u)-1)^2}
\neanl \fl && \quad \quad
-{\cal G}_{3 v} \frac{\sqrt{1-e_\phi ^2}  \sin (u) \left(\left(e_\phi ^2-4\right) \cos (2 u)-7 e_\phi ^2+12
   e_\phi  \cos (u)-2\right)}{2 (e_\phi  \cos (u)-1)^3}
\biggr\}
\,.
\end{eqnarray}
Using the chain rule once more one gets an expression for the angular velocity via $\dot{\phi}(u) = (\diffl{\phi}{v})\, (\diffl{v}{u})\, (\diffl{u}{t})$, symbolically,
\begin{eqnarray}
\label{Eq::dot_phi_u_short}
\fl
\dot{ \phi}(u)&=&
\frac{\Phi}{P}
\frac{\sqrt{1 - e_\phi^2}}{(1 - e_\phi \cos u)} \times
\neanl \fl &&
\Biggl\{
1
+ {\cal G}_{2 v} \, \frac{ 
\left(
3 e_\phi ^2-4 e_\phi  \cos(u) - (e_\phi ^2-2 ) \cos (2 u)
\right)}
{(e_\phi  \cos (u)-1)^2}
\neanl \fl &&
+ {\cal G}_{3 v}\, \frac{
 \left(30 e_\phi ^3-45 e_\phi ^2 \cos (u)-18
   \left(e_\phi ^2-2\right) e_\phi  \cos (2 u)+3 \left(3 e_\phi ^2-4\right) \cos (3 u)\right)}{4 (e_\phi 
   \cos (u)-1)^3} 
\Biggr\}
\times
\neanl \fl &&
\Biggl \{
1-{e_t} \cos u  
+ {\cal F}_v \frac{\sqrt{1-e_\phi ^2} (e_\phi -\cos u  )}{(1-e_\phi  \cos u  )^2}
+ {\cal F}_{v-u}\left[ \frac{\sqrt{1-e_\phi ^2}}{1-e_\phi  \cos u  }-1\right]
\Biggr \}^{-1}
\,.
\end{eqnarray}
Again, $P$ can easily be computed with the help of the already known definition $n \equiv 2\,\pi/P$ and \Eref{Eq::n}.
\section{Conclusions}
In this paper we presented a quasi-Keplerian parameterisation for compact binaries with spin
and arbitrary mass ratio.
We assumed that the spins are aligned or anti-aligned with the orbital angular momentum and restricted
ourselves to the leading-order  spin-spin and next-to-leading order spin-orbit, as well as to 2\pN\
point particle contributions.

The conservation of alignment for all times holds if the alignment is assumed at the
initial instant of time. It turned out that the effects of the spins do not destroy the polynomial structure
of the integrals for both the angular and the radial variables, for which the standard routine
is valid \cite{Schafer:Wex:1993,Damour:Gopakumar:Iyer:2004,Konigsdorffer:Gopakumar:2005}
and enabled us to give a fully analytic prescription for the orbital elements in terms of the
binding energy, the mass ratio and the magnitudes of the angular momenta.

Furthermore, in contrast to the literature where mostly the emphasis was put on the consistent \pN\
accurate presentation of the phasing, we provided \pN\ extended formulae for the radiation
polarisations in analytic form as well. These were derived from the results of \cite{Kidder:1995,Damour:Gopakumar:Iyer:2004} due to the currently highest available order in spin.

We are aware that there is a missing term linear in spin at 2\pN\ order in the wave amplitude.
Blanchet et al. \cite{Blanchet:Buonanno:Faye:2006} provided the current and mass multipole moments
that are necessary to compute the far-zone fluxes resulting from the next-to-leading order spin-orbit
terms in the acceleration, but the wave amplitude at this order was not given. This missing spin-orbit part at
2\pN\ will be given in a forthcoming publication. We justify this decision by stating that there is
a number of relatively complicated terms of higher order due to the transformation from harmonic to
ADM coordinates. To this order, the coordinate transformation contains next-to-leading order spin-orbit
terms which will result in lengthy expressions in the radiation field. The difficulty of computing
the 2\pN\ amplitude itself becomes clear when we keep in mind the errata of reference \cite{Blanchet:Buonanno:Faye:2006}.

An outstanding question is the stability of the spin configurations under purely conservative dynamics. 
If we assume that the spins have tiny differences in their directions, it is interesting to know if the 
enclosed angles will grow secularly or will oscillate in an unknown manner. This will be task of a further
investigation, as well as the inclusion of additional higher order spin Hamiltonians. Aspects of the time
evolution of the misalignment of spins due to the radiation reaction were already discussed by Kidder in \cite{Kidder:1995}. 
%

Another task to be tackled is the effect of radiation reaction to the orbital elements. It is possible to
include the conservative contributions of the spin to the orbital motion into the equations of the far-zone
energy and angular momentum flux expressions.
The goal is an equation of motion for the orbital elements to be obtained in an adiabatic approach.

\ack
We thank Jan Steinhoff and Steven Hergt for many useful discussions.
This work is partly funded by the Deutsche Forschungsgemeinschaft (DFG) through
SFB/TR7 ``Gravitationswellenastronomie'' and the Research Training School GRK 1523 ``Quanten- und Gravitationsfelder''
and by the Deutsches Zentrum f\"ur Luft- und Raumfahrt (DLR) through ``LISA Germany''. 
An anonymous referee's helpful comments and suggestions for improvements
are thankfully acknowledged.
\vspace{2cm}
\bibliographystyle{utphys} 

\begin{thebibliography}{10}

\bibitem{Rowan:Hough:2000}
S.~Rowan and J.~Hough, ``Gravitational wave detection by interferometry (ground
  and space),'' {\em Living Rev. Relativity} {\bf 3} (2000)  3.
  \url{http://www.livingreviews.org/lrr-2000-3}.

\bibitem{Faye:Blanchet:Buonanno:2006}
G.~Faye, L.~Blanchet, and A.~Buonanno, ``{H}igher-order spin effects in the
  dynamics of compact binaries. {I}. {E}quations of motion,''
  \href{http://dx.doi.org/10.1103/PhysRevD.74.104033}{{\em Phys. Rev. D} {\bf
  74} (2006)  104033},
\href{http://arxiv.org/abs/gr-qc/0605139}{{\tt arXiv:gr-qc/0605139}}.

\bibitem{Blanchet:Buonanno:Faye:2006}
L.~Blanchet, A.~Buonanno, and G.~Faye, ``{H}igher-order spin effects in the
  dynamics of compact binaries. {II}. {R}adiation field,''
  \href{http://dx.doi.org/10.1103/PhysRevD.74.104034}{{\em Phys. Rev. D} {\bf
  74} (2006)  104034},
\href{http://arxiv.org/abs/gr-qc/0605140}{{\tt arXiv:gr-qc/0605140}}.

\bibitem{Blanchet:Buonanno:Faye:2006:err}
L.~Blanchet, A.~Buonanno, and G.~Faye, ``Erratum: {H}igher-order spin effects
  in the dynamics of compact binaries. {II}. {R}adiation field,''
  \href{http://dx.doi.org/10.1103/PhysRevD.75.049903}{{\em Phys. Rev. D} {\bf
  75} (2007)  049903(E)}.

\bibitem{Blanchet:Buonanno:Faye:2006:err:2}
L.~Blanchet, A.~Buonanno, and G.~Faye, ``{E}rratum: {H}igher-order spin effects
  in the dynamics of compact binaries. {II}. {R}adiation field,''
  \href{http://dx.doi.org/10.1103/PhysRevD.81.089901}{{\em Phys. Rev. D} {\bf
  81} (2010)  089901(E)}.

\bibitem{Arun:Buonanno:Faye:Ochsner:2009}
K.~G. Arun, A.~Buonanno, G.~Faye, and E.~Ochsner, ``{H}igher-order spin effects
  in the amplitude and phase of gravitational waveforms emitted by inspiraling
  compact binaries: {R}eady-to-use gravitational waveforms,''
  \href{http://dx.doi.org/10.1103/PhysRevD.79.104023}{{\em Phys. Rev. D} {\bf
  79} (2009)  104023}, \href{http://arxiv.org/abs/0810.5336}{{\tt
  arXiv:0810.5336 [gr-qc]}}.

\bibitem{Hannam:Husa:Brugmann:Gopakumar:2008}
M.~Hannam, S.~Husa, B.~Br{\"u}gmann, and A.~Gopakumar, ``{C}omparison between
  numerical-relativity and post-{N}ewtonian waveforms from spinning binaries:
  The orbital hang-up case,''
  \href{http://dx.doi.org/10.1103/PhysRevD.78.104007}{{\em Phys. Rev. D} {\bf
  78} (2008)  104007}, \href{http://arxiv.org/abs/0712.3787}{{\tt
  arXiv:0712.3787 [gr-qc]}}.

\bibitem{Jaranowski:Schafer:1997}
P.~Jaranowski and G.~Sch{\"a}fer, ``Radiative 3.5 post-{N}ewtonian {ADM}
  {H}amiltonian for many-body point-mass systems,''
\href{http://dx.doi.org/10.1103/PhysRevD.55.4712}{{\em Phys. Rev. D} {\bf 55}
  (1997)  4712--4722}.

\bibitem{Blanchet:Faye:Iyer:Joguet:2002}
L.~Blanchet, G.~Faye, B.~R. Iyer, and B.~Joguet, ``{G}ravitational-wave
  inspiral of compact binary systems to 7/2 post-{N}ewtonian order,''
  \href{http://dx.doi.org/10.1103/PhysRevD.65.061501}{{\em Phys. Rev. D} {\bf
  65} (2002)  061501(R)}, \href{http://arxiv.org/abs/gr-qc/0105099}{{\tt
  arXiv:gr-qc/0105099}}.

\bibitem{Blanchet:Faye:Iyer:Joguet:2002:err}
L.~Blanchet, G.~Faye, B.~R. Iyer, and B.~Joguet, ``{E}rratum:
  {G}ravitational-wave inspiral of compact binary systems to 7/2
  post-{N}ewtonian order,''
  \href{http://dx.doi.org/10.1103/PhysRevD.71.129902}{{\em Phys. Rev. D} {\bf
  71} (2005)  129902(E)}.

\bibitem{Ajith:Babak:Chen:others:2008}
P.~Ajith, S.~Babak, Y.~Chen, M.~Hewitson, B.~Krishnan, A.~M. Sintes, J.~T.
  Whelan, B.~Br{\"u}gmann, P.~Diener, N.~Dorband, J.~Gonzalez, M.~Hannam,
  S.~Husa, D.~Pollney, L.~Rezzolla, L.~Santamar{\'i}a, U.~Sperhake, and
  J.~Thornburg, ``{T}emplate bank for gravitational waveforms from coalescing
  binary black holes: {N}onspinning binaries,''
  \href{http://dx.doi.org/10.1103/PhysRevD.77.104017}{{\em Phys. Rev. D} {\bf
  77} (2008)  104017}, \href{http://arxiv.org/abs/0710.2335}{{\tt
  arXiv:0710.2335 [gr-qc]}}.

\bibitem{Ajith:Babak:Chen:others:2008:err}
P.~Ajith, S.~Babak, Y.~Chen, M.~Hewitson, B.~Krishnan, A.~M. Sintes, J.~T.
  Whelan, B.~Br{\"u}gmann, P.~Diener, N.~Dorband, J.~Gonzalez, M.~Hannam,
  S.~Husa, D.~Pollney, L.~Rezzolla, L.~Santamar{\'i}a, U.~Sperhake, and
  J.~Thornburg, ``Erratum: {T}emplate bank for gravitational waveforms from
  coalescing binary black holes: {N}onspinning binaries,''
  \href{http://dx.doi.org/10.1103/PhysRevD.79.129901}{{\em Phys. Rev. D} {\bf
  79} (2009)  129901(E)}.

\bibitem{Mathisson:1937}
M.~Mathisson, ``{N}eue {M}echanik materieller {S}ysteme,''
{\em Acta Phys. Pol.} {\bf 6} (1937)  163--200.

\bibitem{Papapetrou:1951}
A.~Papapetrou, ``{S}pinning test-particles in general relativity. {I},''
\href{http://dx.doi.org/10.1098/rspa.1951.0200}{{\em Proc. R. Soc. A} {\bf 209}
  (1951)  248--258}.

\bibitem{Corinaldesi:Papapetrou:1951}
E.~Corinaldesi and A.~Papapetrou, ``{S}pinning test-particles in general
  relativity. {II},''
\href{http://dx.doi.org/10.1098/rspa.1951.0201}{{\em Proc. R. Soc. A} {\bf 209}
  (1951)  259--268}.

\bibitem{Barker:OConnell:1970}
B.~M. Barker and R.~F. O'Connell, ``{D}erivation of the equations of motion of
  a gyroscope from the quantum theory of gravitation,''
  \href{http://dx.doi.org/10.1103/PhysRevD.2.1428}{{\em Phys. Rev. D} {\bf 2}
  (1970)  1428--1435}.

\bibitem{Barker:OConnell:1975}
B.~M. Barker and R.~F. O'Connell, ``Gravitational two-body problem with
  arbitrary masses, spins, and quadrupole moments,''
\href{http://dx.doi.org/10.1103/PhysRevD.12.329}{{\em Phys. Rev. D} {\bf 12}
  (1975)  329--335}.

\bibitem{DEath:1975}
P.~D. D'Eath, ``Interaction of two black holes in the slow-motion limit,''
  \href{http://dx.doi.org/10.1103/PhysRevD.12.2183}{{\em Phys. Rev. D} {\bf 12}
  (1975)  2183--2199}.

\bibitem{Barker:OConnell:1979}
B.~M. Barker and R.~F. O'Connell, ``The gravitational interaction: Spin,
  rotation, and quantum effects---a review,''
  \href{http://dx.doi.org/10.1007/BF00756587}{{\em Gen. Relativ. Gravit.} {\bf
  11} (1979)  149--175}.

\bibitem{Apostolatos:1995}
T.~A. Apostolatos, ``{S}earch templates for gravitational waves from
  precessing, inspiraling binaries,''
  \href{http://dx.doi.org/10.1103/PhysRevD.52.605}{{\em Phys. Rev. D} {\bf 52}
  (1995)  605--620}.

\bibitem{Kidder:Will:Wiseman:1993}
L.~E. Kidder, C.~M. Will, and A.~G. Wiseman, ``{S}pin effects in the inspiral
  of coalescing compact binaries,''
  \href{http://dx.doi.org/10.1103/PhysRevD.47.R4183}{{\em Phys. Rev. D} {\bf
  47} (1993)  R4183--R4187},
\href{http://arxiv.org/abs/gr-qc/9211025}{{\tt arXiv:gr-qc/9211025}}.

\bibitem{Damour:Deruelle:1985}
T.~Damour and N.~Deruelle, ``{G}eneral relativistic celestial mechanics of
  binary systems. {I}. the post-{N}ewtonian motion.,'' {\em Ann. Inst. H.
  Poincar{\'e} A} {\bf 43} (1985)  107--132.

\bibitem{Schafer:Wex:1993}
G.~Sch{\"a}fer and N.~Wex, ``{S}econd post-{N}ewtonian motion of compact
  binaries,'' \href{http://dx.doi.org/doi:10.1016/0375-9601(93)90758-R}{{\em
  Phys. Lett. A} {\bf 174} (1993)  196--205}.

\bibitem{Schafer:Wex:1993:err}
G.~Sch{\"a}fer and N.~Wex, ``Erratum: {S}econd post-{N}ewtonian motion of
  compact binaries,''
  \href{http://dx.doi.org/doi:10.1016/0375-9601(93)90980-E}{{\em Phys. Lett. A}
  {\bf 177} (1993)  461(E)}.

\bibitem{Memmesheimer:Gopakumar:Schafer:2004}
R.-M. Memmesheimer, A.~Gopakumar, and G.~Sch{\"a}fer, ``{T}hird
  post-{N}ewtonian accurate generalized quasi-{K}eplerian parametrization for
  compact binaries in eccentric orbits,''
  \href{http://dx.doi.org/10.1103/PhysRevD.70.104011}{{\em Phys. Rev. D} {\bf
  70} (2004)  104011},
\href{http://arxiv.org/abs/gr-qc/0407049}{{\tt arXiv:gr-qc/0407049}}.

\bibitem{Reisswig:Husa:Rezzolla:Dorband:Pollney:Seiler:2009}
C.~Reisswig, S.~Husa, L.~Rezzolla, E.~N. Dorband, D.~Pollney, and J.~Seiler,
  ``{G}ravitational-wave detectability of equal-mass black-hole binaries with
  aligned spins,'' \href{http://dx.doi.org/10.1103/PhysRevD.80.124026}{{\em
  Phys. Rev. D} {\bf 80} (2009)  124026},
  \href{http://arxiv.org/abs/0907.0462}{{\tt arXiv:0907.0462 [gr-qc]}}.

\bibitem{Bogdanovic:Reynolds:Miller:2007}
T.~Bogdanovi{\'c}, C.~S. Reynolds, and M.~C. Miller, ``{A}lignment of the spins
  of supermassive black holes prior to coalescence,''
  \href{http://dx.doi.org/10.1086/518769}{{\em ApJ} {\bf 661} (2007)
  L147--L150}, \href{http://arxiv.org/abs/gr-qc/0703054}{{\tt
  arXiv:gr-qc/0703054}}.

\bibitem{Steinhoff:Hergt:Schafer:2008:2}
J.~Steinhoff, S.~Hergt, and G.~Sch{\"a}fer, ``Next-to-leading order
  gravitational spin(1)-spin(2) dynamics in {H}amiltonian form,''
  \href{http://dx.doi.org/10.1103/PhysRevD.77.081501}{{\em Phys. Rev. D} {\bf
  77} (2008)  081501(R)},
\href{http://arxiv.org/abs/0712.1716}{{\tt arXiv:0712.1716 [gr-qc]}}.

\bibitem{Hergt:Steinhoff:Schafer:2010:1}
S.~Hergt, J.~Steinhoff, and G.~Sch{\"a}fer, ``The reduced {H}amiltonian for
  next-to-leading-order spin-squared dynamics of general compact binaries,''
  \href{http://dx.doi.org/10.1088/0264-9381/27/13/135007}{{\em Class. Quant.
  Grav.} {\bf 27} (2010)  135007},
\href{http://arxiv.org/abs/1002.2093}{{\tt arXiv:1002.2093 [gr-qc]}}.

\bibitem{MartinGarcia:2008}
J.~M. Mart{\'i}n-Garc{\'i}a, ``x{P}erm: fast index canonicalization for tensor
  computer algebra,'' \href{http://dx.doi.org/10.1016/j.cpc.2008.05.009}{{\em
  Comp. Phys. Commun.} {\bf 179} (2008)  597--603},
  \href{http://arxiv.org/abs/0803.0862}{{\tt arXiv:0803.0862 [cs.SC]}}.

\bibitem{MartinGarcia:2002}
J.~M. Mart{\'i}n-Garc{\'i}a, {\em x{A}ct: Efficient Tensor Computer Algebra}.
\newblock \url{http://www.xact.es/}.

\bibitem{Jaranowski:Schafer:1998}
P.~Jaranowski and G.~Sch{\"a}fer, ``{T}hird post-{N}ewtonian higher order {ADM}
  {H}amilton dynamics for two-body point-mass systems,''
  \href{http://dx.doi.org/10.1103/PhysRevD.57.7274}{{\em Phys. Rev. D} {\bf 57}
  (1998)  7274--7291}, \href{http://arxiv.org/abs/gr-qc/9712075}{{\tt
  arXiv:gr-qc/9712075}}.

\bibitem{Damour:Jaranowski:Schafer:2001}
T.~Damour, P.~Jaranowski, and G.~Sch{\"a}fer, ``{D}imensional regularization of
  the gravitational interaction of point masses,''
  \href{http://dx.doi.org/10.1016/S0370-2693(01)00642-6}{{\em Phys. Lett. B}
  {\bf 513} (2001)  147--155},
\href{http://arxiv.org/abs/gr-qc/0105038}{{\tt arXiv:gr-qc/0105038}}.

\bibitem{Poisson:1998}
E.~Poisson, ``Gravitational waves from inspiraling compact binaries: {T}he
  quadrupole-moment term,''
  \href{http://dx.doi.org/10.1103/PhysRevD.57.5287}{{\em Phys. Rev. D} {\bf 57}
  (1998)  5287--5290},
\href{http://arxiv.org/abs/gr-qc/9709032}{{\tt arXiv:gr-qc/9709032}}.

\bibitem{Damour:Jaranowski:Schafer:2008:1}
T.~Damour, P.~Jaranowski, and G.~Sch{\"a}fer, ``{H}amiltonian of two spinning
  compact bodies with next-to-leading order gravitational spin-orbit
  coupling,'' \href{http://dx.doi.org/10.1103/PhysRevD.77.064032}{{\em Phys.
  Rev. D} {\bf 77} (2008)  064032},
\href{http://arxiv.org/abs/0711.1048}{{\tt arXiv:0711.1048 [gr-qc]}}.

\bibitem{Steinhoff:Schafer:Hergt:2008}
J.~Steinhoff, G.~Sch{\"a}fer, and S.~Hergt, ``{ADM} canonical formalism for
  gravitating spinning objects,''
  \href{http://dx.doi.org/10.1103/PhysRevD.77.104018}{{\em Phys. Rev. D} {\bf
  77} (2008)  104018},
\href{http://arxiv.org/abs/0805.3136}{{\tt arXiv:0805.3136 [gr-qc]}}.

\bibitem{Levi:2008}
M.~Levi, ``Next to leading order gravitational spin1-spin2 coupling with
  {K}aluza-{K}lein reduction,''
\href{http://arxiv.org/abs/0802.1508}{{\tt arXiv:0802.1508 [gr-qc]}}.

\bibitem{Thorne:Hartle:1985}
K.~S. Thorne and J.~B. Hartle, ``Laws of motion and precession for black holes
  and other bodies,''
\href{http://dx.doi.org/10.1103/PhysRevD.31.1815}{{\em Phys. Rev. D} {\bf 31}
  (1985)  1815--1837}.

\bibitem{Hergt:Schafer:2008:2}
S.~Hergt and G.~Sch{\"a}fer, ``Higher-order-in-spin interaction {H}amiltonians
  for binary black holes from source terms of {K}err geometry in approximate
  {ADM} coordinates,'' \href{http://dx.doi.org/10.1103/PhysRevD.77.104001}{{\em
  Phys. Rev. D} {\bf 77} (2008)  104001},
\href{http://arxiv.org/abs/0712.1515}{{\tt arXiv:0712.1515 [gr-qc]}}.

\bibitem{Hergt:Schafer:2008}
S.~Hergt and G.~Sch{\"a}fer, ``Higher-order-in-spin interaction {H}amiltonians
  for binary black holes from {P}oincar{\'e} invariance,''
  \href{http://dx.doi.org/10.1103/PhysRevD.78.124004}{{\em Phys. Rev. D} {\bf
  78} (2008)  124004},
\href{http://arxiv.org/abs/0809.2208}{{\tt arXiv:0809.2208 [gr-qc]}}.

\bibitem{Steinhoff:Schafer:2009:2}
J.~Steinhoff and G.~Sch{\"a}fer, ``Canonical formulation of self-gravitating
  spinning-object systems,''
  \href{http://dx.doi.org/10.1209/0295-5075/87/50004}{{\em Europhys. Lett.}
  {\bf 87} (2009)  50004},
\href{http://arxiv.org/abs/0907.1967}{{\tt arXiv:0907.1967 [gr-qc]}}.

\bibitem{Steinhoff:Wang:2009}
J.~Steinhoff and H.~Wang, ``Canonical formulation of gravitating spinning
  objects at 3.5 post-{N}ewtonian order,''
  \href{http://dx.doi.org/10.1103/PhysRevD.81.024022}{{\em Phys. Rev. D} {\bf
  81} (2010)  024022},
\href{http://arxiv.org/abs/0910.1008}{{\tt arXiv:0910.1008 [gr-qc]}}.

\bibitem{Tulczyjew:1959}
W.~M. Tulczyjew, ``{M}otion of multipole particles in general relativity
  theory,'' {\em Acta Phys. Pol.} {\bf 18} (1959)  393--409.

\bibitem{Dixon:1979}
W.~G. Dixon, ``Extended bodies in general relativity: {T}heir description and
  motion,'' in {\em Proceedings of the International School of Physics Enrico
  Fermi LXVII}, J.~Ehlers, ed., pp.~156--219.
\newblock North Holland, Amsterdam, 1979.

\bibitem{Steinhoff:Puetzfeld:2009}
J.~Steinhoff and D.~Puetzfeld, ``{M}ultipolar equations of motion for extended
  test bodies in general relativity,''
  \href{http://dx.doi.org/10.1103/PhysRevD.81.044019}{{\em Phys. Rev. D} {\bf
  81} (2010)  044019},
\href{http://arxiv.org/abs/0909.3756}{{\tt arXiv:0909.3756 [gr-qc]}}.

\bibitem{Laarakkers:Poisson:1999}
W.~G. Laarakkers and E.~Poisson, ``{Q}uadrupole moments of rotating neutron
  stars,'' \href{http://dx.doi.org/10.1086/306732}{{\em Astrophys. J.} {\bf
  512} (1999)  282--287}, \href{http://arxiv.org/abs/gr-qc/9709033}{{\tt
  arXiv:gr-qc/9709033}}.

\bibitem{Levin:2000}
J.~Levin, ``{G}ravity waves, chaos, and spinning compact binaries,''
  \href{http://dx.doi.org/10.1103/PhysRevLett.84.3515}{{\em Phys. Rev. Lett.}
  {\bf 84} (2000)  3515--3518}, \href{http://arxiv.org/abs/gr-qc/9910040}{{\tt
  arXiv:gr-qc/9910040}}.

\bibitem{Sohr:2009}
C.~Sohr, ``{C}haos in gravitativ-gebundenen {B}in{\"a}rsystemen mit {S}pin.,''
  diploma thesis, Friedrich-Schiller-Universit{\"a}t Jena, 2009.
\newblock unpublished.

\bibitem{Dirac:1964}
P.~A.~M. Dirac, {\em Lectures on Quantum Mechanics}.
\newblock Yeshiva University Press, New York, 1964.

\bibitem{Goldstein:1981}
H.~Goldstein, {\em Classical Mechanics}.
\newblock Addison-Wiley Publishing Company, 2nd~ed., 1981.

\bibitem{Colwell:1993}
P.~Colwell, {\em {S}olving {K}epler's equation over three centuries}.
\newblock Willman-Bell, Inc., Richmond, VA 23235, 1993.

\bibitem{Mikoczi:Vasuth:Gergely:2005}
B.~Mik{\'o}czi, M.~Vas{\'u}th, and L.~{\'A}. Gergely, ``{S}elf-interaction spin
  effects in inspiralling compact binaries,''
  \href{http://dx.doi.org/10.1103/PhysRevD.71.124043}{{\em Phys. Rev. D} {\bf
  71} (2005)  124043}, \href{http://arxiv.org/abs/astro-ph/0504538}{{\tt
  arXiv:astro-ph/0504538}}.

\bibitem{Majar:Vasuth:2006}
J.~Maj{\'a}r and M.~Vas{\'u}th, ``{G}ravitational waveforms from a
  {L}ense-{T}hirring system,''
  \href{http://dx.doi.org/10.1103/PhysRevD.74.124007}{{\em Phys. Rev. D} {\bf
  74} (2006)  124007}, \href{http://arxiv.org/abs/gr-qc/0611105}{{\tt
  arXiv:gr-qc/0611105}}.

\bibitem{Vasuth:Majar:2007}
M.~Vas{\'u}th and J.~Maj{\'a}r, ``{G}ravitational waveforms for finite mass
  binaries,'' \href{http://dx.doi.org/10.1142/S0217751X07036488}{{\em Int. J.
  Mod. Phys. A} {\bf 22} (2007)  2405--2414},
  \href{http://arxiv.org/abs/0705.3481}{{\tt arXiv:0705.3481 [gr-qc]}}.

\bibitem{Majar:Vasuth:2008}
J.~Maj{\'a}r and M.~Vas{\'u}th, ``Gravitational waveforms for spinning compact
  binaries,'' \href{http://dx.doi.org/10.1103/PhysRevD.77.104005}{{\em Phys.
  Rev. D} {\bf 77} (2008)  104005},
\href{http://arxiv.org/abs/0806.2273}{{\tt arXiv:0806.2273 [gr-qc]}}.

\bibitem{Gergely:2009}
L.~{\'A}. Gergely, ``{S}pinning compact binary inspiral: {I}ndependent
  variables and dynamically preserved spin configurations,''
  \href{http://arxiv.org/abs/0912.0459}{{\tt arXiv:0912.0459}}.

\bibitem{Tessmer:2009}
M.~Tessmer, ``{G}ravitational waveforms from unequal-mass binaries with
  arbitrary spins under leading order spin-orbit coupling,''
  \href{http://dx.doi.org/10.1103/PhysRevD.80.124034}{{\em Phys. Rev. D} {\bf
  80} (2009)  124034},
\href{http://arxiv.org/abs/0910.5931}{{\tt arXiv:0910.5931 [gr-qc]}}.

\bibitem{Keresztes:Mikoczi:Gergely:2005}
Z.~Keresztes, B.~Mik{\'o}czi, and L.~{\'A}. Gergely, ``The {K}epler equation
  for inspiralling compact binaries,''
  \href{http://dx.doi.org/10.1103/PhysRevD.72.104022}{{\em Phys. Rev. D} {\bf
  72} (2005)  104022},
\href{http://arxiv.org/abs/astro-ph/0510602}{{\tt arXiv:astro-ph/0510602}}.

\bibitem{Damour:Gopakumar:Iyer:2004}
T.~Damour, A.~Gopakumar, and B.~R. Iyer, ``{P}hasing of gravitational waves
  from inspiralling eccentric binaries,''
  \href{http://dx.doi.org/10.1103/PhysRevD.70.064028}{{\em Phys. Rev. D} {\bf
  70} (2004)  064028}, \href{http://arxiv.org/abs/gr-qc/0404128}{{\tt
  arXiv:gr-qc/0404128}}.

\bibitem{Konigsdorffer:Gopakumar:2006}
C.~K{\"o}nigsd{\"o}rffer and A.~Gopakumar, ``{P}hasing of gravitational waves
  from inspiralling eccentric binaries at the third-and-a-half post-{N}ewtonian
  order,'' \href{http://dx.doi.org/10.1103/PhysRevD.73.124012}{{\em Phys. Rev.
  D} {\bf 73} (2006)  124012}, \href{http://arxiv.org/abs/gr-qc/0603056}{{\tt
  arXiv:gr-qc/0603056}}.

\bibitem{Gergely:Perjes:Vasuth:1998}
L.~{\'A}. Gergely, Z.~I. Perj{\'e}s, and M.~Vas{\'u}th, ``Spin effects in
  gravitational radiation backreaction. {III}: {C}ompact binaries with two
  spinning components,''
  \href{http://dx.doi.org/10.1103/PhysRevD.58.124001}{{\em Phys. Rev. D} {\bf
  58} (1998)  124001},
\href{http://arxiv.org/abs/gr-qc/9808063}{{\tt arXiv:gr-qc/9808063}}.

\bibitem{Kidder:1995}
L.~E. Kidder, ``Coalescing binary systems of compact objects to
  (post)$^{5/2}$-{N}ewtonian order. {V}. {S}pin effects,''
  \href{http://dx.doi.org/10.1103/PhysRevD.52.821}{{\em Phys. Rev. D} {\bf 52}
  (1995)  821--847},
\href{http://arxiv.org/abs/gr-qc/9506022}{{\tt arXiv:gr-qc/9506022}}.

\bibitem{Gopakumar:Iyer:1997}
A.~Gopakumar and B.~R. Iyer, ``Gravitational waves from inspiraling compact
  binaries: Angular momentum flux, evolution of the orbital elements, and the
  waveform to the second post-newtonian order,''
  \href{http://dx.doi.org/10.1103/PhysRevD.56.7708}{{\em Phys. Rev. D} {\bf 56}
  (1997)  7708--7731}, \href{http://arxiv.org/abs/gr-qc/0110100}{{\tt
  arXiv:gr-qc/0110100}}.

\bibitem{Damour:Jaranowski:Schafer:2001:2}
T.~Damour, P.~Jaranowski, and G.~Sch{\"a}fer, ``{E}quivalence between the
  {ADM}-{H}amiltonian and the harmonic-coordinates approaches to the third
  post-{N}ewtonian dynamics of compact binaries,''
  \href{http://dx.doi.org/10.1103/PhysRevD.63.044021}{{\em Phys. Rev. D} {\bf
  63} (2001)  044021}, \href{http://arxiv.org/abs/gr-qc/0010040}{{\tt
  arXiv:gr-qc/0010040}}.

\bibitem{Konigsdorffer:Gopakumar:2005}
C.~K{\"o}nigsd{\"o}rffer and A.~Gopakumar, ``{P}ost-{N}ewtonian accurate
  parametric solution to the dynamics of spinning compact binaries in eccentric
  orbits: {T}he leading order spin-orbit interaction,''
  \href{http://dx.doi.org/10.1103/PhysRevD.71.024039}{{\em Phys. Rev. D} {\bf
  71} (2005)  024039}, \href{http://arxiv.org/abs/gr-qc/0501011}{{\tt
  arXiv:gr-qc/0501011}}.

\bibitem{Damour:Schafer:1985}
T.~Damour and G.~Sch{\"a}fer, ``{L}agrangians for n point masses at the second
  post-{N}ewtonian approximation of general relativity,''
  \href{http://dx.doi.org/10.1007/BF00773685}{{\em Gen. Relativ. Gravit.} {\bf
  17} (1985)  879--905}.

\end{thebibliography}
\providecommand{\href}[2]{#2}\begingroup\raggedright\endgroup

\appendix{
\section{Integrals}\label{AppA}
For the sake of completeness we give the results of the definite integrals $I_n$ and $I_n'$  for different $n$:
\begin{eqnarray}
\fl I^\prime_0 &=&\frac{\pi  (\sm+\sp)}{(\sm \sp)^{3/2}},\eanl
\fl I^\prime_1 &=&\frac{2 \pi }{\sqrt{\sm \sp}}, \eanl
\fl I^\prime_2 &=& 2 \pi,  \eanl
\fl I^\prime_3 &=&\pi  (\sm+\sp), \eanl
\fl I^\prime_4 &=&\frac{1}{4} \pi  \left(3 \sm^2+2 \sm \sp+3 \sp^2\right), \eanl
\fl I^\prime_5 &=&\frac{1}{8} \pi  (\sm+\sp) \left(5 \sm^2-2 \sm \sp+5 \sp^2\right).
\end{eqnarray}
The more complicated integrals with boundary $s$ in terms of 
$u$, 
$\tilde{v}$, 
$e_r$ and 
$a_r$ are given by
\begin{eqnarray}
\fl I_0 &=& a_r^2 \sqrt{1-e_r^2}\,(u - \sin u), \eanl
\fl I_1 &=& a_r \sqrt{1-e_r^2}\,u, \eanl
\fl I_2 &=& \tilde{v}, \eanl
\fl I_3 &=& \frac{\tilde{v}+e_r \sin \left(\tilde{v}\right)}{a_r \left(1-e_r^2\right)}, \eanl
\fl I_4 &=& \frac{2 (2 + e_r^2) \tilde{v}+8 e_r \sin \tilde{v}+e_r^2 \sin \left(2 \tilde{v}\right)}{4 a_r^2 \left(1-e_r^2\right){}^2}, \eanl
\fl I_5 &=& \frac{6 \left(2 + 3 e_r^2\right) \tilde{v} +9 e_r \left(4 + e_r^2\right)\sin \tilde{v}+9 e_r^2 \sin \left(2 \tilde{v}\right)+e_r^3 \sin \left(3 \tilde{v}\right)}{12 a_r^3 \left(1-e_r^2\right){}^3}. 
\end{eqnarray}

\section{Coordinate transformation from ADM to harmonic}\label{AppB}
From section IV of  \cite{Damour:Jaranowski:Schafer:2008:1} and from \cite{Damour:Jaranowski:Schafer:2001:2},
we collect the contributions for the coordinate transformation from ADM to harmonic coordinates for spinning
compact binaries, including LO effects of spin-orbit interaction and 2\pN\ PP contributions.
Let  $\vct{Y}_a$ label the harmonic position of the $a$-th particle as a function of the ADM positions
$\vct{x}_b$, momenta $\vct{p}_b$ and spins $\vct{S}_b$. Then, to 2\pN\ order, the transformation reads
{\em in their notation}

\begin{eqnarray}
\fl
\label{eq11}
\vct{Y}_a(\vct{x}_b,\vct{ p}_b)
= \vct{ x}_a
+ \epso^2 \, \vct{ Y}^{\rm{SO}}_a(\vct{ x}_b,\vct{ p}_b, \vct{ S}_b)
+ \epso^4 \, \vct{ Y}^{\rm{2PN}}_a(\vct{ x}_b,\vct{ p}_b)
\end{eqnarray}
with
\begin{eqnarray}
\fl
\vct{ Y}^{\rm{SO}}_a(\vct{ x}_b,\vct{ p}_b, \vct{ S}_b)
&=&
\frac{\vct{ S}_a \times \vct{ p}_a}{2\, m_a^2} \label{eq:x_trafo_spin} \,,
\eanl
 \label{eq12}
\fl
\vct{ Y}^{\rm{2PN}}_1(\vct{ x}_a,\vct{ p}_a)
&=& G m_2 \Bigg\{ \left[ \frac{5}{8} \frac{\vct{p}_{2}^2}{m_2^2} - \frac{1}{8} 
\frac{\scp{\vnunit}{\vct{p}_{\rm{2}}}^2}{m_2^2}
+ \frac{G m_1}{r_{12}} \left(\frac{7}{4} + \frac{1}{4} \frac{m_2}{m_1}\right) 
\right]
\vct{ n}_{12}
\nonumber\\[2ex]&&
\phantom{G m_2 \Bigg\{}
+ \frac{1}{2} \frac{\scp{\vnunit}{\vct{p}_{{\rm 2}}} }{m_2} \frac{\vct{ p}_1}{m_1}
-  \frac{7}{4} \frac{\scp{\vnunit}{\vct{p}_{{\rm 2}}} }{m_2} \frac{\vct{ p}_2}{m_2} \Bigg\}
\,,
\end{eqnarray}
where $\vct{ Y}^{\rm{2PN}}_2(\vct{ x}_a,\vct{ p}_a)$ is simply obtained by exchanging the particle indices (1 $\leftrightarrow$ 2).
We find it very important to mention some of the rules to obtain the relative separation vector with the scaling introduced in this paper.
The above equations are {\em not} given in relative coordinates. Thus, we scale
every $\vct{S}_a$ with $m_a^2$. Next, we subtract $\vct{ Y}_1$ from $\vct{Y}_2$,
setting $\vct{p}_2=-\vct{p}_1=-\vct{p}$
for the centre-of-mass frame and scale $\vct{p}$ with $\mu$ as in \Eref{eq::pscale} to get a dimensionless momentum.
Finally, we divide the obtained separation vector with $G\,m$ and obtain the separation in terms of the linear momentum
and the ADM spin momenta.

There is an additional transformation at 2PN which relates the ADM time with the time in
harmonic coordinates \cite{Damour:Schafer:1985}, which reads
\begin{equation}
t^{\rm ADM} = t^{\rm h} + \epsilon^4 \, \eta \, \scp{\vnunit}{\vct{v}}
\,
\end{equation}
where we removed the scales and used only dimensionless terms.

The harmonic velocity is obtained by plugging the harmonic positions in the Poisson brackets with the
total Hamiltonian and adding the internal derivation of the ADM time with respect to the harmonic time,
\begin{eqnarray}
 \vvelocity			&=&	\left[ \vct{ {x}}^{		}, \HAM{}{ADM} \right] \,, \\
 \vvelocity^{\rm{hcov}}&=&	\left[ \vct{ {x}}^{\rm hcov}, \HAM{}{ADM} \right]\,
\left( \frac{\rmd t^{\rm ADM}}{\rmd t^{\rm hcov}} \right)
 \,,
\end{eqnarray}
The linear momentum $\vct{p}$ can then be expressed in terms of the velocity perturbatively.
It is important to express $\vct{p}$ in terms of the ADM velocity first and then to plug it into the expression for $\vct{ v}^{\rm hcov}$
afterwards.
To 2\pN\ order, the radial separation, the velocity and the unit normal vector, $\rel^{\rm{harm}}$, $\vct{ v}^{\rm harm}$ and  $\vnunit^{\rm{harm}}$
transform due to
\begin{eqnarray}
\fl
\vct{ x}^{\rm harm} &=&
\vct{x}
+\frac{1}{2} \epso ^2\spino  \eta  
   \crp{\vct{S}}{\vct{v}}
+\epso ^4 \left\{\frac{12 \eta +1}{4 {\rel} }\vnunit
-\frac{1}{8} \eta  \left(\vnunit
  \left({\dot \rel}  ^2-5  {\vct{v}^2} \right)+18 {\dot \rel}  
   \vct{v}  \right)\right\}
\,,
\eanl  
\fl
\vct{v}^{\rm{harm}} &=&
\vct{v}
- \epso ^2 \, \spino \,\frac{\eta}{2 {\rel}  ^2}
\crp{\vct{S}}{\vnunit}
\neanl \fl && 
\frac{1 }{8 \, c^4}
\Biggl\{
\vnunit_{}
 \left(\eta 
   \left[
\frac{
 3 
\dot{r}^2_{}-7 
v^{2}_{}
}{r_{}}
-\frac{38
}{r^2_{}}\right]
-\frac{4}{r^2_{}}\right)
\dot{r}_{}
+\vct{v}
   \left(\eta  
\left[
\frac{9 \dot{r}^2_{}-5 v^{2}_{} }{r_{}}
+\frac{34}{r^2_{}}
\right]
+\frac{2}{r^2_{}}\right)
\Biggr\}
\,.
\eanl 
\fl \rel^{\rm{harm}} &=&
{\rel}
-\frac{1}{2}  \epso^2 \spino   \eta \SPAT{\vct{S}}{\vnunit}{\vct{v}}
+\epso ^4 \left\{\frac{1}{8} \eta  \left(5  {\vct{v}^2} -19
   {\dot \rel}  ^2\right)+\frac{3 \eta +\frac{1}{4}}{{\rel}  }\right\}
\,,
\eanl  
\fl  \vnunit^ {\rm harm} &=&
\vnunit
+\epso ^2  \, \spino  \,   \frac{\eta}{2 {\rel}  }\, \left\{\vnunit \SPAT{\vct{S}}{\vnunit}{\vct{v}}
+ \crp{\vct{S}}{\vct{v}}
\right\}
+\epso ^4 \frac{9 {\dot \rel}   \eta }{4 {\rel}  }  \left\{{\dot \rel}   \vnunit  -\vct{v}  \right\}
\,,
\end{eqnarray}
where every quantity on the right hand side is written in ADM coordinates. Note that
\begin{eqnarray}
 \vnunit \SPAT{\vct{S}}{\vnunit}{\vct{v}} + \crp{\vct{S}}{\vct{v}} & = & ({\bf 1} - \vnunit\otimes\vnunit) \crp{\vct{S}}{\vct{v}}
\end{eqnarray}
is the part of $\crp{\vct{S}}{\vct{v}}$ which is orthogonal to $\vnunit$.
}

\end{document}